\documentclass[journal]{IEEEtran}

\IEEEoverridecommandlockouts                           
\usepackage[ruled,vlined,linesnumbered]{algorithm2e}
\usepackage{amssymb,amsmath,amsfonts,amsthm}
\usepackage{bbm}
\usepackage{bm}
\usepackage{balance}
\usepackage{cite}
\usepackage[T1]{fontenc}
\usepackage[latin9]{inputenc}
\usepackage{graphicx}
\usepackage{hyperref} 
\usepackage{soul} 
\usepackage[usenames,dvipsnames]{xcolor}
\usepackage{verbatim}
\usepackage{xspace}
\usepackage{cases}
\usepackage{tikz}
\usetikzlibrary{plotmarks}

\newcommand{\T}{^{\mbox{\tiny T}}}

\def\tr{\mathop{\mathrm{tr}}}

\newcommand{\E}{\ensuremath{\mathbb{E}}}
 
\newcommand{\PR}{\mathbb{P}} 
\newcommand{\norm}[1]{\left\lVert#1\right\rVert}

\def\CoIL{\mathop{\mathrm{CoIL}}}

\newenvironment{list4}{
\begin{list}{$\bullet$}{%
	\setlength{\itemsep}{0.05cm}
	\setlength{\labelsep}{0.2cm}
	\setlength{\labelwidth}{0.3cm}
	\setlength{\parsep}{0in} 
	\setlength{\parskip}{0in}
	\setlength{\topsep}{0in} 
	\setlength{\partopsep}{0in}
	\setlength{\leftmargin}{0.17in}}}
{\end{list}}

\newtheorem{remark}{Remark}
\newtheorem{theorem}{Theorem}
\newtheorem{lemma}{Lemma}

\newtheorem{example}{\bfseries Example}

\newtheorem{definition}{Definition}

\newcommand{\RW}[1]{\textcolor{black}{{#1}}}

\hypersetup{
colorlinks = true,
linkcolor = [rgb]{0,0,1},
anchorcolor = [rgb]{0,0,1},
citecolor = [rgb]{0.9,0.5,0},
filecolor = [rgb]{0,0,1},
pagecolor = [rgb]{1,1,0},
urlcolor = [rgb]{0.9,0.5,0},
bookmarks,
bookmarksopen = true,
bookmarksnumbered = true,
breaklinks = true,
linktocpage,
pagebackref,
colorlinks = true,
linkcolor = [rgb]{0.2,0.6,0.2},
urlcolor  = [rgb]{0,0,1},
citecolor = [rgb]{0.9,0.5,0},
anchorcolor = [rgb]{0.2,0.6,0.2},
hyperindex = true,
hyperfigures
}
\newcommand{\overbar}[1]{\mkern 1.5mu\overline{\mkern-1.5mu#1\mkern-1.5mu}\mkern 1.5mu}
\DeclareMathOperator*{\argmax}{arg\,max}
\DeclareMathOperator*{\argmin}{arg\,min}

\usepackage{pgfplots}
\pgfplotsset{compat=newest}
\usetikzlibrary{plotmarks}
\usetikzlibrary{arrows.meta}
\usepgfplotslibrary{patchplots}
\usepackage{grffile}

\usepackage{pgfplots}
\pgfplotsset{compat=newest}
\usetikzlibrary{plotmarks}
\usetikzlibrary{arrows.meta}
\usepgfplotslibrary{patchplots}
\usepackage{grffile}
\usepackage{amsmath}

\pgfplotsset{plot coordinates/math parser=false}
\newlength\fheight 
\newlength\fwidth
\begin{document}

\title{\huge Distributed Channel Access for Control Over Unknown Memoryless Communication Channels}

\author{Tahmoores Farjam, Henk Wymeersch, and Themistoklis Charalambous% <-this % stops a space
\thanks{T. Farjam and T. Charalambous are with the Department of Electrical Engineering and Automation, School of Electrical Engineering, Aalto University, Espoo, Finland (e-mails: name.surname@aalto.fi).}
\thanks{H. Wymeersch is with the Department of Electrical Engineering, Chalmers University of Technology, G\"{o}teborg, Sweden (e-mail: henkw@chalmers.se).}
\thanks{A preliminary version of the results appeared in \cite{Farjam:2019a}. Here, we consider a different scenario where the sensors are equipped with a local Kalman filter and readjust our approach accordingly. Furthermore, we establish the stability conditions and also propose a method to ensure that adopting indexing policies results in collision-free channel access.
}
}

\maketitle

% ==================================================
%
%
% ABSTRACT
%
%
% ==================================================
\begin{abstract}
	We consider the distributed channel access problem for a system	consisting of multiple control subsystems that close their loop over a shared wireless network. We propose a distributed method for providing deterministic channel access without requiring explicit information exchange between the subsystems. This is achieved by utilizing timers for prioritizing channel access with respect to a local cost which we derive by transforming the control objective cost to a form that allows its local computation. This property is then exploited for developing our distributed deterministic channel access scheme. A framework to verify the stability of the system under the resulting scheme is then proposed. Next, we consider a practical scenario in which the channel statistics are unknown. We propose learning algorithms for learning the parameters of imperfect communication links for estimating the channel quality and, hence, define the local cost as a function of this estimation and control performance. We establish that our learning approach results in collision-free channel access. The behavior of the overall system is exemplified via a proof-of-concept illustrative example, and the efficacy of this mechanism is evaluated for large-scale networks via simulations.
\end{abstract}


% ==================================================
%
%
% INTRODUCTION
%
%
% ==================================================
\section{Introduction}\label{sec:intro}

Modern control environments involve various control loops, each containing numerous spatially distributed smart devices with sensing, actuating and computing capabilities. In Wireless Networked Control Systems (WNCSs), the wireless medium is used for information exchange between these devices, which greatly facilitates fast and easy deployments at lower installation and maintenance cost than their wired counterparts. Nevertheless, incorporating wireless channels in control loops introduces new challenges that need to be addressed for realizing the full potential of WNCSs \cite{Zhang:2017}.

One of the key challenges of WNCSs is the unreliable nature of wireless communication which leads to packet dropouts. In contrast to classical control settings, where the communication is assumed to be ideal, the effect of such imperfections can no longer be neglected. In the seminal work \cite{Schenato:2007}, the impact of lossy communication links in control loops was thoroughly investigated. It was shown that the certainty equivalence principle holds as long as the adopted network protocols guarantee packet acknowledgements/negative-acknow\-ledge\-ments (ACK/NACKs). In addition, when only the link between the sensor and the controller is unreliable, the closed-loop performance can be investigated by considering the impact of packet dropouts on the corresponding Kalman filter. As a result, the prominent results obtained for sensor scheduling in remote estimation for various channel models, communication and energy constraints can readily be adopted in respective WNCS settings; see, e.g., \cite{Wu:2014, Wu:2013, Nourian:2014, Han:2017, Quevedo:2013, Huang:2007, Leong:2012, Knorn:2017}. %

Typically, WNCSs contain several control subsystems which share the same communication resources. Due to the limited available bandwidth, the wireless devices need to coordinate for sharing the scarce and unreliable communication resources efficiently to accomplish the control tasks and with a good performance. This has led to a surge of research on the design of effective resource allocation schemes in the last decade; see, for example, the recent survey \cite{Park:2018} and references therein. It has been shown that finding the optimal schedule for multiple subsystems that have access to multiple lossy channels requires solving a mixed-integer quadratic program which is computationally infeasible for large systems \cite{Zanon:2018, HernandezLerma:1996}. In the presence of a central coordinator in the network, this can be overcome by employing priority-based resource allocation schemes, which determine the priorities dynamically with respect to a finite-horizon criterion. Try-once-discard (TOD) is one of the most well-known schemes of this type which, at each frame, allocates the channels to the subsystems with the largest discrepancy between the true and estimated state values \cite{Walsh:2002}. For the linear quadratic Gaussian (LQG) control problem, the \emph{value of information} (VoI) contained in the sensors' current observations for the network was proposed as the priority measure in \cite{Molin:2015, Molin:2019}. In a similar context, the contribution of the loss of data packet for a controller on the increase of the quadratic cost of the entire system was labeled as the \emph{cost of information loss} (CoIL) in \cite{Charalambous:2017}. For decoupled systems, it was shown that minimizing the linear quadratic cost is equivalent to prioritizing channel access with respect to CoIL. In case of sensors with limited energy budget, the energy expenditure can also be included in the objective or as a constraint for determining transmission priorities as proposed in \cite{Gatsis:2014, Gatsis:2015, Leong:2017}.

In some scenarios, a central coordinator is non-existent, thus requiring the subsystems to coordinate in a distributed manner for accessing the channels. The channel access in such settings is often provided by implementing random medium access schemes, which are incapable of taking the control performance into account, e.g., ALOHA. However, a limited number of novel deterministic solutions have emerged for resolving contention over perfect channels showing promising results \cite{Mamduhi:2017, Farjam:2018}. In this paper, we consider the LQG scenario and address how the scarce and unreliable communication resources can be allocated in a distributed manner by proposing a priority-based channel access scheme. This is achieved by adopting a variant of the timer-based mechanism for CoIL (TBCoIL), which was initially proposed in \cite{Farjam:2018} for contention resolution over ideal communication channels. The proposed variant of this mechanism allows for distributed channel access in settings where sensors have access to multiple wireless channels subject to independent and identically distributed (i.i.d.) packet dropouts. Moreover, this mechanism leads to control-aware channel access as long as the timer values are defined accordingly. To this end, we utilize the concept of CoIL and show that the optimal timer setup requires knowledge of the rate of packet dropouts. In practice, however, the sufficient statistics of the probability distributions according to which the packet dropouts happen are unknown. To enable implementation in such practical scenarios, we propose a method for learning the essential channel parameters online and in a control-aware manner.
% The main feature of this mechanism for WNCSs is that the timer values can be defined such that control-aware channel access. This mechanism can be implemented in WNCSs  granting channel access in a control-aware manner by adopting a . Based on the concept of CoIL, we then derive the optimal timer setup which can be implemented if the statistics of the packet dropping channels are known \emph{a priori}.

Applying learning methods in control problems has a long history being mostly centered around learning the unknown system dynamics by reinforcement learning; see \cite{Recht:2019} for a thorough literature review. However, only a limited number of works consider learning methods in the context of scheduling for WNCSs. In \cite{Wu:2019}, the problem of sensor scheduling with communication rate constraints over a single channel is addressed. After proving the threshold-like structure of the optimal scheduling policy, iterative algorithms are designed to obtain the optimal solution without knowing the packet dropout rate. In a similar context, the relationship between the sample complexity and stability margin of a system over an unknown memoryless channel was investigated in \cite{Gatsis:2021}. Most recently, a multi-armed bandit (MAB) approach was proposed for near-optimal resource allocation~\cite{Wang:2020}. In the case of multiple stable systems having access to multiple known identical channels, the optimal scheduling problem was solved by deriving the Whittle's index leading to promising performance at low computational cost. In this paper, we also utilize the celebrated results obtained for MAB problems. However, our setup poses unique challenges since it includes unstable subsystems, which need to coordinate in a distributed manner for accessing multiple available channels. Furthermore, each wireless link can have a distinct packet dropout rate which is unknown. To overcome these, we cast our problem as a MAB one and propose a novel distributed solution, which also takes the control performance into account. 

The main contributions of this paper can be summarized as follows:
\begin{list4}
\item We propose \RW{a distributed} channel access method for WNCSs with multiple unreliable communication links. This is achieved by extending the application of TBCoIL to multi-channel wireless networks. We then utilize the concept of CoIL to formulate the channel access problem for minimizing the stage cost and show that utilizing timers for solving this problem in distributed manner requires \emph{a priori} knowledge of the packet dropout rates.
\item We propose a framework for verifying the mean square stability of the system under the proposed channel access scheme. By modeling the packet arrival sequence as a Markov chain and investigating its stationary distribution, we derive the sufficient conditions that guarantee stability. Furthermore, we illustrate how the proposed framework can be utilized to determine stability in practical settings where analytical expressions for the stationary distribution cannot be derived.
%\TF{We derive the mean square stability conditions by modeling the packet arrival sequence under the proposed channel access scheme as a Markov chain and investigating its stationary distribution. Furthermore, we demonstrate how the proposed framework can be utilized for verifying stability in practical settings.}  
\item We additionally consider the practical scenario in which the channel statistics are unknown \emph{a priori}. We first demonstrate how the well-known indexing policies developed for single-player MABs can be employed in timers to solve multi-player MABs in a distributed and collision-free manner. Next, we cast the channel access problem as a multi-player MAB and introduce time-varying weights for scaling the indices with respect to the control performance. The resulting control-dependent indices are then utilized in timers to solve the channel access problem in a distributed and control-aware manner.
%For the case of unknown channels, we cast the channel selection problem as a MAB and devise a novel policy for minimizing the finite-horizon cost criterion. We first utilize the well-known indexing policies developed for single-player MABs in multi-player settings and propose a method for enabling distributed collision-free implementation. Then, time-varying weights are introduced for scaling the indices with respect to the control performance. Finally, the policy is enforced in a distributed fashion by using the proposed variant of TBCoIL.
\end{list4}

The remainder of this paper is organized as follows: the necessary preliminaries and system model are provided in Section~\ref{sec:model}. The proposed distributed channel access mechanism for channels with known statistics is described in Section~\ref{sec:mechanism} and stability conditions are derived. In Section~\ref{sec:MAB}, we use the MAB approach for learning the unknown channel parameters and propose a novel method for concurrent distributed channel access and learning. In Section~\ref{sec:numericals}, we evaluate the effectiveness of the proposed method by numerical simulations. Finally, we draw conclusions and discuss future directions in Section~\ref{sec:conclusion}.

\emph{Notation:} 
%Vectors and matrices are denoted by lowercase and uppercase letters, respectively. 
$\mathbb{Z}_{\geq 0}$ ($\mathbb{Z}_{>0}$) denotes the set of nonnegative (positive) integers. The transpose, inverse, and trace of a square matrix $X$ are denoted by $X^{T}$, $X^{-1}$, and $\tr(X)$, respectively, while the notation $X\succeq 0$ ($X\succ 0$) means that matrix $X$ is positive semi-definite (definite). $\mathbb{S}_+^n$ is the set of $n$ by $n$ positive semi-definite matrices and the $n$ by $n$ identity matrix is represented by $I_n$. $\mathbb{E}\{\cdot\}$ represents the expectation of its argument and $\mathbb{P}\{\cdot\}$ denotes the probability of an event. $f^n(\cdot)$ is the $n$-fold composition of $f(\cdot)$, with the convention that $f^0(X)=X$. The Euclidean norm of a vector $x$ is denoted by $\norm{x}$ and $\sigma_{\max}(X)$ denotes the spectral radius of a matrix $X$. Finally, the cardinality of a set $\mathcal{X}$ is denoted by $\vert \mathcal{X} \vert$.

% ==================================================
%
%
% SYSTEM MODEL
%
%
% ==================================================
\section{System Model and Preliminaries}\label{sec:model}
\begin{figure}[t]
	\centering
	\includegraphics[width=\columnwidth]{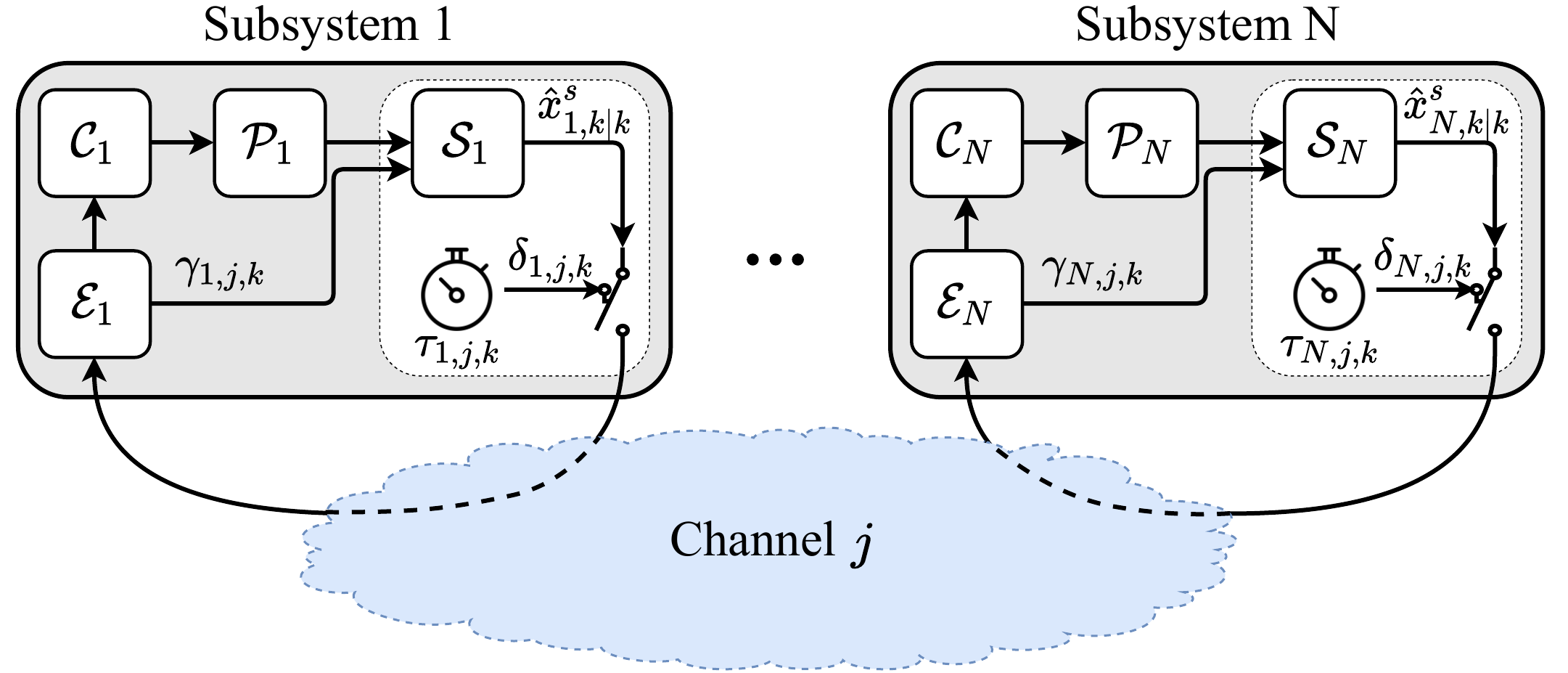}
	\vspace{-0.25cm}
	\caption{Example of the WNCS layout where $N$ subsystems compete to access a shared wireless channel $j$. $\mathcal{P}_i$ represents the plant of subsystem $i\in\{1,\ldots,N\}$, with $\mathcal{S}_i$, $\mathcal{E}_i$, and $\mathcal{C}_i$ being its sensor, estimator and controller, respectively. Note that the timer is embedded in the smart sensor and determines whether $\hat{x}_{i,k|k}^s$ is transmitted from $\mathcal{S}_i$ to $\mathcal{E}_i$.}
	\label{fig:diagram}
\end{figure}

\RW{The schematic diagram of the WNCS under consideration is depicted in Fig.~\ref{fig:diagram}. Each subsystem consists of an unstable dynamical process, a dedicated local controller, smart sensor, and estimator. We assume the actuators are collocated with the controllers, but the sensors need to transmit their data over a capacity-constrained time-slotted network. We consider the scenario in which the subsystems exchange no explicit information and they coordinate for channel access in a distributed manner\footnote{\RW{The term \emph{decentralized control} is often used to describe scenarios in which determining the control inputs requires no explicit information exchange between subsystems. In this work, we use the term \emph{distributed} to describe the channel access problem in accordance with the WNCSs literature \cite{Park:2018}.}}. The effects of state quantization and transmission delays are considered negligible and are thus ignored henceforth.}
%Due to the limited bandwidth, in each time slot, only a limited number of sensors can transmit their data packets successfully. 

%%%%%%%%%%%%%%%%%%%%%%%%%%%%%
\subsection{Local Processes and Measurements}
The process dynamics are assumed to be unstable and modeled by the following linear time-invariant (LTI) stochastic process: 
\begin{subequations}\label{eq:process}
\begin{align}
	x_{i,k+1} &= A_{i}x_{i,k}+B_{i}u_{i,k}+w_{i,k},\\
	y_{i,k} &= C_{i}x_{i,k}+v_{i,k},
\end{align}
\end{subequations}
where $x_{i,k} \in \mathbb{R}^{n_{i}}$, $y_{i,k} \in \mathbb{R}^{p_{i}}$, and $u_{i,k} \in \mathbb{R}^{m_{i}}$ are the states, outputs measured by the sensor, and inputs of subsystem $i$ at time step $k$, respectively. $A_{i}$, $B_{i}$ and $C_{i}$ are the system, input and output matrices of appropriate dimensions and to avoid trivial cases, we assume that $\sigma_{\max}(A_{i})>1$ for all $i$. Moreover, $w_{i,k}$ and $v_{i,k}$ are the uncorrelated zero-mean Gaussian disturbance and measurement noise, respectively, with respective covariances $W_i\succeq 0$, and $V_{i}\succ 0$. The initial state $x_{i,0}$ is also a Gaussian random variable with mean $\bar{x}_{i,0}$ and covariance $X_{i,0}\succeq 0$, which is independent of $w_{i,k}$ and $v_{i,k}$, i.e., $\E\{x_{i,0}w_{i,k}^T\} =\E\{x_{i,0}v_{i,k}^T\}=0$ for all $k$.

Each sensor is assumed to have enough computational power for pre-processing the measurement data. Similar to the typical configuration considered in remote estimation, we consider the scenario in which the sensors compute the state estimate and transmit that rather than the raw measurement. The resulting estimator outperforms the one based on raw (unprocessed) measurements since it results in smaller error covariance while offering tighter stability conditions \cite{Gupta:2007, Schenato:2008}. Let $\mathcal{I}_{i,k}^s=\{y_{i,0},\ldots,y_{i,k}\}$ denote the available information set at the sensor of subsystem $i$ at time $k$ and define
\begin{align*}
\hat{x}_{i,k|k-1}^s&\triangleq \E\{x_{i,k} | \mathcal{I}^s_{i,k-1}\}, \qquad \hat{x}_{i,k|k}^s\triangleq \E\{x_{i,k} | \mathcal{I}^s_{i,k}\}, \\
P_{i,k|k-1}^s&\triangleq\E\{(x_{i,k}-\hat{x}_{i,k|k-1}^s)(x_{i,k}-\hat{x}_{i,k|k-1}^s)\T | \mathcal{I}_{i,k-1}^s\},\\
P_{i,k|k}^s&\triangleq\E\{(x_{i,k}-\hat{x}_{i,k|k}^s)(x_{i,k}-\hat{x}_{i,k|k}^s)\T | \mathcal{I}_{i,k}^s\}.
\end{align*}
Due to availability of complete observation history, the sensor can compute the minimum mean square error (MMSE) estimate of the state by running a local Kalman filter which computes $\hat{x}_{i,k|k}^s$ to be transmitted to the corresponding estimator recursively by
\begin{subequations}
\begin{align*}
	\hat{x}_{i,k|k-1}^s &= A_{i}\hat{x}_{i,k-1|k-1}^s+B_iu_{i,k-1},\\
	P_{i,{k|k-1}}^s &= h_i(P_{i,{k-1|k-1}}^s),  \\%\label{eq:pricov}\\
	K_{i,k} &= P_{i,{k|k-1}}^s C_i\T( C_iP_{i,{k|k-1}}^s C_i\T + V_i )^{-1},\\
	\hat{x}_{i,k|k}^s &= \hat{x}_{i,k|k-1}^s + K_{i,k} (y_{i,k}-C_i\hat{x}_{i,k|k-1}^s),  \\%\label{eq:posx}\\
	P_{i,{k|k}}^s &= g_i\circ h_i(P_{i,{k-1|k-1}}^s),%\label{eq:poscov},
\end{align*}
\end{subequations}
where the functions $h,\,g: \mathbb{S}_+^n \to \mathbb{S}_+^n$ are defined as
\begin{align*}
h_i(X) &\triangleq A_i X A_i\T + W_i,\\
g_i(X) &\triangleq X - X C_i\T (C_i X C_i\T + V_i)^{-1} C_i X.
\end{align*} 
\RW{By assuming that the pair $(A_i,C_i)$ is observable and $(A_i,W_i^{1/2})$ is controllable, $g_i\circ h_i (X)=X$ has a unique positive semi-definite solution. We denote this solution by $\overbar{P}_i$ which represents the steady-state error covariance of subsystem $i$. In such settings, it is commonly assumed that the Kalman filter has entered steady state since the \emph{a posteriori} error covariance converges to $\overbar{P}_i$ exponentially fast for any initial conditions \cite{BrianD.O.Anderson:2012}. Since all the parameters for evaluating $g_i\circ h_i (X)=X$ are known, we initiate the filter from $P_{i,{0|-1}}^s=\overbar{P}_{i}$ and $\hat{x}_{i,0|-1}^s=0$ to ensure that it is already in steady state.}
%Denote $\overbar{P}_i$ as the steady-state error covariance of subsystem $i$, i.e., $\overbar{P}_i$ is the unique positive semi-definite solution to $g_i\circ h_i (X)=X$, \TF{which always exists based on the assumption that the pair $(A_i,C_i)$ is observable, $(A_i,W_i^{1/2})$ is controllable}. Without loss of generality, we assume that the Kalman filter has entered steady state and $P_{i,{k|k}}^s=\overbar{P}_i$ since the \emph{a posteriori} error covariance converges to $\overbar{P}_i$ exponentially fast in this setting \cite{BrianD.O.Anderson:2012}.
%with the initial conditions $\hat{x}_{i,0|-1}^s=0$ and $P_{i,{0|-1}}^s=X_{0,i}$. 

%%%%%%%%%%%%%%%%%%%%%%%%%%%%%
\subsection{Imperfect Communication}	

Let $\mathcal{N}$ and $\mathcal{M}$ denote the index set of subsystems and available channels, respectively, with $\vert\mathcal{N}\vert=N$ and $\vert\mathcal{M}\vert=M$. The problem of scheduling typically arises when the shared communication resources are limited, i.e., $M<N$, as is the case considered here. Let the decision variable $\delta_{i,j,k} \in \{0,1\}$ represent whether subsystem $i$ transmits on channel $j$ at time step $k$ as follows
\begin{align*}
\delta_{i,j,k} = \begin{cases}
	1, & \hat{x}_{i,k|k}^s\text{ is transmitted on channel }j, \\
	0, & \text{otherwise}.
\end{cases}
\end{align*}
We assume packet ACK/NACKs are guaranteed at each time instant and define another binary variable $\gamma_{i,j,k}$ such that $\gamma_{i,j,k}=1$ corresponds to the event of successful packet reception given that $\delta_{i,j,k}=1$; otherwise, $\gamma_{i,j,k}=0$. Similarly, to represent whether a subsystem $i$ receives the data packet at $k$, regardless of the selected channel, we define an additional binary variable $\theta_{i,k}$ as 
\begin{align*}
\theta_{i,k} = \begin{cases}
	1, & \sum_{j\in\mathcal{M}}\gamma_{i,j,k}=1, \\
	0, & \text{otherwise}.
\end{cases}
\end{align*}

To ensure collision-free channel access, we impose a constraint on the network such that a channel can only be accessed by one subsystem at a given time
\begin{align} \label{eq:constraint1}
\sum_{i\in\mathcal{N}}\delta_{i,j,k} \leq 1, \quad \forall j, \forall k.
\end{align} 
Moreover, since it is assumed that one slot is enough to convey all the information from the sensor to the estimator at each time slot $k$, each subsystem can use one channel at most, i.e.,
\begin{align} \label{eq:constraint2}
\sum_{j\in\mathcal{M}}\delta_{i,j,k} \leq 1, \quad \forall i, \forall k.
\end{align}

The communication channels in this work are non-ideal and thus prone to packet dropouts due to the effects of phenomena such as multipath fading, shadowing, interference, etc. This unreliability can be taken into consideration by modeling the packet dropouts over each channel as i.i.d. Bernoulli random sequences, and consequently the probability of successful transmission satisfies a Bernoulli distribution with mean $q_{i,j}\in(0,1]$. Using the introduced decision variables, the probability of successful packet delivery over a wireless link is given by
\begin{align} \label{eq:Psucc}
\PR\{\gamma_{i,j,k} = 1|\delta_{i,j,k}=1\} = q_{i,j}.
\end{align}
%%%%%%%%%%%%%%%%%%%%%%%%%%%%%%
\subsection{Control and Estimation}

In this work, the standard quadratic cost over the infinite horizon is chosen as the performance metric. This cost is defined as	
\begin{align}\label{eq:cost}
J_{0:\infty} = \lim\limits_{K \to \infty}\frac{1}{K}\mathbb{E}\left\{\sum_{k=0}^{K-1}\sum_{i\in\mathcal{N}}\left(x\T_{i,k}Q_ix_{i,k}+u\T_{i,k}R_i u_{i,k}\right)\right\},
\end{align}
where $Q_i\succeq$ and $R_i\succ 0$ are weighting matrices of appropriate dimensions. As long as the channel access decisions are independent of the control inputs, the certainty equivalence principle holds and the optimal controller for minimizing this cost is linear and obtained by 
%	\begin{align} \label{eq:input}
$u_{i,k} = L_{i,\infty}\hat{x}_{k|k}$,
%	\end{align}
where $L_{i,\infty}$ is the optimal feedback gain given by 
\begin{align*}
L_{i,\infty} = -(B\T_i\Pi_{i,\infty} B_i + R_i)^{-1}B\T_i\Pi_{i,\infty} A_i,
\end{align*}
where $\Pi_{i,\infty}$ is the positive semi-definite solution of discrete-time algebraic Riccati equation (DARE)
\begin{align*}
\Pi_{i,\infty}=A_i\T \Pi_{i,\infty} A_i + Q_i - L_{i,\infty}\T(B_i\T \Pi_{i,\infty} B_i+R_i)L_{i,\infty},
\end{align*}
which always exists due to perfect actuation links and assuming that the pairs $(A_i,B_i)$ and $(A_i,Q_i^{1/2})$ are controllable and observable, respectively \cite{Chen:1995}.
The local estimator computes the state estimate, denoted by $\hat{x}_{i,k|k}$, based on the information received from the sensor. Let $t_{i,k}\triangleq\min\{\kappa\geq0:\theta_{i,{k-\kappa}}=1\}$ denote the time elapsed since the most recent successful transmission. Based on its locally available information, i.e., $\mathcal{I}_{i,k}=\{\theta_{i,1},\hat{x}_{i,1|1}^s\theta_{i,1},\ldots,\theta_{i,k},\hat{x}_{i,k|k}^s\theta_{i,k}\}$, the estimator computes
\begin{align*}
\hat{x}_{i,k|k}&= (A_i+B_iL_{i,\infty})^{t_{i,k}}\hat{x}_{i,k-t_{i,k}|k-t_{i,k}}^s,\\
P_{i,k|k}&= h_i^{t_{i,k}}(\overbar{P}_i),
\end{align*}
where $P_{i,k|k}\triangleq\E\{{e}_{i,k|k}{e}_{i,k|k}\T|\mathcal{I}_{i,k}\}$ denotes the error covariance at the estimator with the error being defined as ${e}_{i,k|k}\triangleq x_{i,k}-\hat{x}_{i,k|k}$. Moreover, $\overbar{P}_i$ is the steady-state \emph{a posteriori} error covariance at the corresponding sensor. This estimation architecture is equivalent to the optimal estimator that one would obtain if all observations up to time $k-t_{i,k}$ were successfully delivered \cite{Schenato:2008}.

%%%%%%%%%%%%%%%%%%%%%%%%%%%%%
\subsection{Cost of Information Loss (CoIL)}

The concept of CoIL was introduced in \cite{Charalambous:2017} to quantify the additional cost incurred due to the loss of information. Let $\mathcal{F}_k\subseteq\mathcal{N}$ denote the set of subsystems that transmit their data packet at $k$ and $\overbar{\mathcal{F}}_k\triangleq \mathcal{N} \setminus\mathcal{F}_k$. Furthermore, we define $E_{i,k}^0$ as the cost of subsystem $i$ in case it does not receive any data at $k$; similarly, $E_{i,k}^1$ is the cost when this subsystem receives its data packet. Let $j_i:\mathcal{F}_k\to\mathcal{M}$ denote the index $j$ with $\delta_{i,j,k}=1$ for $i\in \mathcal{F}_k$. At the beginning of time slot $k$, the expected value of the stage cost, denoted by $J_k$, can be written as
\begin{align} \label{eq:stCost}
\E&\{J_k|{\mathcal{I}^{k-1}},\mathcal{F}_k,\mathcal{Q}\} \nonumber \\
&= \sum_{i\in\overbar{\mathcal{F}}_k}E_{i,k}^0 + \sum_{i\in \mathcal{F}_k}\left(E_{i,k}^0(1-q_{i,j_i})+E_{i,k}^1q_{i,j_i}\right) \nonumber \\
&= \sum_{i\in \mathcal{N}}E_{i,k}^0 + \sum_{i\in \mathcal{F}_k}\left(E_{i,k}^1 - E_{i,k}^0\right)q_{i,j_i} \nonumber \\
&= \sum_{i\in\mathcal{N}}E_{i,k}^0 - \sum_{i\in \mathcal{F}_k}{\CoIL}_{i,k}q_{i,j_i},
\end{align}
where $\mathcal{I}^{k-1}\triangleq\cup_{i\in\mathcal{N}} \mathcal{I}_{i,k-1}$ is the available information at the beginning of time slot $k$, $\mathcal{Q}\triangleq\{q_{i,j}:\forall i\in\mathcal{N},\forall j\in\mathcal{M}\}$ contains the probability of successful transmission over each wireless link, and $\CoIL_{i,k}\triangleq E_{i,k}^0 - E_{i,k}^1$ denotes the \emph{cost of information loss}. Consequently, minimizing the expected cost is equivalent to finding $\mathcal{F}_k$ such that the last term is maximized. ${\CoIL}_{i,k}$ can be construed as the amount a subsystem $i$ increases the entire cost, in case it receives no data packet at $k$.

%%%%%%%%%%%%%%%%%%%%%%%%%%%%%
\subsection{Timer-based Mechanism}
The timer-based mechanism, denoted as TBCoIL, was first proposed in \cite{Farjam:2018} for providing channel access in Networked Control Systems (NCSs). This mechanism is able to provide collision-free distributed channel access in capacity constrained networks with no packet dropouts. TBCoIL is based on the idea of assigning a local timer to each subsystem $i$. At each time step $k$, the value of each timer is calculated by
\begin{align}\label{eq:lambda}
\tau_{i,k} = \frac{\lambda}{m_{i,k}},
\end{align}
where $m_{i,k}$ is a nonzero cost which represents how critical the data packet of subsystem $i$ is at time $k$. By choosing a cost that can be calculated by the local information as $m_{i,k}$, all subsystems can set their timers without requiring any explicit information exchange. Note that $\lambda$ in \eqref{eq:lambda} can be interpreted as a tuning parameter which allows for adjusting the duration of the contention period based on the requirements. By using an identical value for $\lambda$ in all subsystems, the subsystem with the largest cost will have the smallest timer.

At the beginning of each transmission slot, subsystems calculate \eqref{eq:lambda} and start their timer from the obtained value. The timer of the subsystem with the largest cost reduces to zero first and the corresponding subsystem sends a short-duration flag packet on the network which informs the remaining contestants to stop their timers and back off to avoid collisions. Then, this subsystem transmits its data packet for the remaining duration of the slot. As the next transmission slot begins, the timers are reset to newly calculated values and the same procedure is repeated. Note that the timer is embedded in the sensor where the local computations are being done for deciding whether to access the channel as depicted in Fig.~\ref{fig:diagram}. The idea of this mechanism in which the timer is a function of the channel quality \emph{only} is a celebrated result in wireless cooperative networks \cite{Bletsas:2006}.

% =====================================================
%
%
% Method Explanation
%
%
% =====================================================
\section{Distributed Channel Access Mechanism}\label{sec:mechanism}

We first modify TBCoIL to extend its application to the case where multiple imperfect channels are available. We assume that each subsystem is equipped with $M$ independent timers, i.e., a separate timer for each channel. Similar to the original method, the timer values are inversely proportional to the local cost and are determined by
\begin{align}\label{eq:lambda_j}
\tau_{i,j,k} = \frac{\lambda_j}{m_{i,j,k}},
\end{align}
where $\lambda_j$ is a constant specific to channel $j\in\mathcal{M}$ but is identical for all $i$, and the nonzero local cost, denoted by $m_{i,j,k}$, is calculated individually for each channel. For simplicity, we will assume that $\lambda_j$ is the same for all channels, i.e., $\lambda_j=\lambda$ for all $j$.

As the transmission slot begins, subsystems start their timers from \eqref{eq:lambda_j}. The smallest timer corresponds to the largest cost and thus the highest priority. Let $\{i^*,j^*\}=\argmin_{i,j} \{\tau_{i,j,k}\}$ which represent the indices of the smallest timer at $k$. As this timer reaches zero, subsystem $i^*$ transmits a short-duration flag packet on channel $j^*$ immediately, which informs other subsystems to stop their timers for $j^*$ and back off. Simultaneously, $i^*$ stops the rest of its timers, i.e., withdraws from competition for other channels, and transmits its data packet on $j^*$ without collision. Meanwhile, the remaining subsystems compete for the remaining available resources until all $M$ channels have been allocated. Therefore, this mechanism inherently satisfies constraints \eqref{eq:constraint1} and \eqref{eq:constraint2}. Similar to the original method, as the time slot ends, all timers are reset (based on the newly calculated local costs) and the entire procedure is repeated in the next time slot.

\begin{remark}
The idea of assigning multiple timers to subsystems can be realized by assuming that each subsystem is equipped with a single real-time clock. %Although, for ease of explanation, we assume that each subsystem has multiple timers, implementation only requires a single clock. 
The value of an imaginary timer assigned to a specific channel can equivalently be represented by a checkpoint on the elapsed time of the clock from the beginning of the respective time slot. As the clock reaches the first checkpoint, i.e., the smallest timer expires, the corresponding channel is claimed and all the remaining checkpoints are removed, which can be interpreted as withdrawing from competition for the remaining resources. Furthermore, if a flag packet is received on a channel, the corresponding checkpoint is neglected which is equivalent to backing off for avoiding packet collision.
\end{remark}

\subsection{Timer Setup}
The main challenge for implementing the proposed channel access mechanism is quantifying the local cost such that it corresponds to control performance while, to avoid explicit information exchange, it is a function of the local information only. We first start by breaking down the quadratic cost of the system to identify the components that are affected by channel access decisions and derive the associated CoIL. As we will show, local information is sufficient for computing CoIL, and we can utilize it in timers for solving the control-aware channel access problem in a distributed fashion.
%The main challenge for implementing the proposed channel access mechanism is the quantification of the local cost, since distributed implementation requires this cost to be a function of local information only. We first start by decomposing the quadratic cost for the entire system and then establish the associated CoIL. Finally, we exploit the characteristics of the obtained result to determine the optimal timer-setup. 
\begin{lemma} \label{lemma:1}
Consider the cost criterion defined in \eqref{eq:cost}. The stage cost at $k$ is given by
\begin{align}\label{eq:stage-cost}
	J_k=\sum_{i\in\mathcal{N}}\tr(\Pi_{i,\infty}W_{i})+\sum_{i\in\mathcal{N}}\tr(\Gamma_{i,\infty}\E\{P_{i,k|k}\}),
\end{align}
where $\Gamma_{i,\infty} = L_{i,\infty}\T(B_i\T \Pi_{i,\infty} B_i + R_i)L_{i,\infty}$.
\end{lemma}

\begin{proof}
Considering a finite horizon, the linear quadratic cost \eqref{eq:cost} can be written as \cite[Lemma 6.1, Ch. 8]{Astrom:2006}
\begin{align*}
{J}_{0:K-1} &{=}\E\left\{\sum_{i\in\mathcal{N}}x_{i,0}\T\Pi_{i,0} x_{i,0}\right\}{+}\E \left\{ \sum_{k=0}^{{K}-1}  \sum_{i\in\mathcal{N}} w_{i,k}\T \Pi_{i,k} w_{i,k} \right\} \notag\\
	&\quad + \E \left\{ \sum_{k=0}^{{K}-1}\sum_{i\in\mathcal{N}} (x_{i,k} - \hat{x}_{i,k|k})\T \Gamma_{i,k}  (x_{i,k} - \hat{x}_{i,k|k}) \right\},
\end{align*}
where $\Pi_{i,k}$ is determined by solving the standard DARE over the finite horizon $K$ and is used accordingly for obtaining the associated matrices $L_{i,k}$ and $\Gamma_{i,k}$. Note that the covariance of the process noise is time-invariant. Hence, using \cite[Lemma 3.3, Ch. 8]{Astrom:2006} yields
\begin{align*}
	{J}_{0:K-1} {=}&\sum_{i\in\mathcal{N}}\tr\left(\Pi_{i,0} (\bar{x}_{i,0}\bar{x}_{i,0}\T {+} X_{i,0})\right){+} \sum_{k=0}^{{K}-1}\sum_{i\in\mathcal{N}} \tr\left(\Pi_{i,k} W_{i} \right) \notag\\
	&+ \sum_{k=0}^{{K}-1}\sum_{i\in\mathcal{N}} \tr(\Gamma_{i,k}\E\{{e}_{i,k|k}{e}_{i,k|k}\T\}) \notag\\
	=&\sum_{i\in\mathcal{N}}\tr\left(\Pi_{i,0} (\bar{x}_{i,0}\bar{x}_{i,0}\T {+} X_{i,0})\right){+} \sum_{k=0}^{{K}-1}\sum_{i\in\mathcal{N}} \tr\left(\Pi_{i,k} W_{i} \right) \notag\\
	&+ \sum_{k=0}^{{K}-1}\sum_{i\in\mathcal{N}} \tr(\Gamma_{i,k}\E\{P_{i,k|k}\}),
\end{align*}
where the first equality results from the zero-mean property of $w_{i,k}$, the definition of ${e}_{i,k|k}\triangleq x_{i,k}-\hat{x}_{i,k|k}$ and recalling that $X_{i,0}$ and $W_i$ are the covariance matrices of the initial state and process disturbance. Moreover, the last equality follows from the law of total expectation and the definition of the error covariance. Therefore, by considering the infinite horizon and using the steady state values $\Gamma_{i,\infty}$ and $\Pi_{i,\infty}$ for all $i$ the stage cost at $k$ is determined by \eqref{eq:stage-cost}.
%By taking the limit for $K\to\infty$ and averaging, the first term approaches zero and the assertion follows.
\end{proof}

We are now ready to derive CoIL by examining how the channel access decisions impact the stage cost in \eqref{eq:stage-cost}.
\begin{lemma} \label{lemma:coil}
The cost of information loss for each subsystem can be formulated as
\begin{align} \label{eq:coil}
	{\CoIL}_{i,k}=\tr\left(\Gamma_{i,\infty}\left[h_i^{t_{i,k-1}+1}(\overbar{P}_i)-\overbar{P}_i\right]\right).
\end{align}
\end{lemma}
\begin{proof}
Following the same procedure as \eqref{eq:stCost} yields %\HW{[Page break in equation]}
\begin{align} \label{eq:J_k}
	&\E\{J_k|\mathcal{I}^{k-1},\mathcal{F}_k,{\mathcal{Q}}\} = \sum_{i\in\mathcal{N}}\tr(\Pi_{i,\infty}W_{i})\notag \\
	& \quad \quad +\sum_{i\in\mathcal{N}}\tr\left(\Gamma_{i,\infty} \E \left\{  P_{i,k|k} |\mathcal{I}_{i,k-1},\mathcal{F}_k, {\mathcal{Q}} \right\} \right)  \notag\\%\displaybreak
	& \quad = \sum_{i\in\mathcal{N}}\tr(\Pi_{i,\infty}W_{i}) +\sum_{i\in\overbar{\mathcal{F}}_k}\tr\left(\Gamma_{i,\infty} h_i^{t_{i,k-1}+1}(\overbar{P}_{i})\right) \notag \\ 
	& \quad \quad +\sum_{i\in \mathcal{F}_k}\tr\left(\Gamma_{i,\infty} \overbar{P}_i\right)q_{i,j_i} \notag\\
	&\quad \quad + \sum_{i\in \mathcal{F}_k}\tr\left(\Gamma_{i,\infty} h_i^{t_{i,k-1}+1}(\overbar{P}_{i})\right)(1-q_{i,j_i}) \notag\\
	& \quad = \sum_{i\in\mathcal{N}}\tr\left(\Pi_{i,\infty}W_{i} + \Gamma_{i,\infty} h_i^{t_{i,k-1}+1}(\overbar{P}_{i})\right) \notag\\
	&\quad \quad - \sum_{i\in \mathcal{F}_k}\tr\left( \Gamma_{i,\infty} \left[h_i^{t_{i,k-1}+1}(\overbar{P}_{i}) - \overbar{P}_i\right] \right)q_{i,j_i}.
\end{align}
Thus, by comparing this result with the definition in \eqref{eq:stCost}, CoIL is obtained by \eqref{eq:coil}.
\end{proof}

Since minimizing \eqref{eq:J_k} is equivalent to maximizing the last term, the optimal resource allocation at time slot $k$ can be formulated as
\begin{align} \label{eq:maximization}
\begin{split}
	\max_{\Delta_{k}} &\quad \sum_{i\in\mathcal{N}}\sum_{j\in\mathcal{M}}{\CoIL}_{i,k}q_{i,j}\delta_{i,j,k}, \\
	\text{subject to}&\quad \eqref{eq:constraint1}, \eqref{eq:constraint2},
\end{split}
\end{align} 	
where CoIL is given by \eqref{eq:coil} and $\Delta_{k}$ is a binary matrix that includes all the optimization variables at time $k$, i.e.,
\begin{align*}
\Delta_{k} \triangleq 
\begin{bmatrix}
	\delta_{1,1,k} & \ldots& \delta_{1,M,k} \\
	\delta_{2,1,k} & \ldots& \delta_{2,M,k} \\
	\vdots & & \vdots \\
	\delta_{N,1,k} & \ldots& \delta_{N,M,k} 
\end{bmatrix}.
\end{align*}
\RW{Note that this optimization problem can be rewritten as a generic \emph{assignment problem} (see \cite{Charalambous:2017}) and solved efficiently in a centralized manner by adopting methods such as the Hungarian method \cite{Kuhn:2005}.} Nevertheless, We aim at solving this problem in a distributed manner. As aforementioned, if the local information is sufficient for determining the cost $m_{i,j,k}$ in \eqref{eq:lambda_j}, implementing the timer-based mechanism ensures that channel access is granted to the subsystems with the highest cost in a distributed fashion. Given that $q_{i,j}$ is known for all $i$ and $j$, as computation of CoIL in \eqref{eq:coil} only requires local information, the product in \eqref{eq:maximization} can be used as $m_{i,j,k}$ in timers, i.e., 
\begin{align} \label{eq:timer}
\tau_{i,j,k} = \frac{\lambda}{{\CoIL}_{i,k}q_{i,j}}.
\end{align}
The first $M$ timers that expire, each for a different channel, determine the transmitting subsystems and the corresponding claimed channels. As a result, this setup provides a distributed solution to \eqref{eq:maximization}. \RW{Since $q_{i,j}$ has Lebesgue measure zero, assuming negligible propagation delays and one-bit flags ensures that channel access is collision-free. Even in homogeneous networks, i.e., network containing subsystems with identical dynamics, timers lead to collision-free channel access since $q_{i,j}$'s are distinct despite (possibly) identical values for ${\CoIL}_{i,k}$.} Note that computation of CoIL only requires knowledge of the system parameters, initial condition $\overbar{P}_{i}$, and the age of the last successfully received packet, i.e., $t_{i,k}$. Therefore, it can be determined independently of the measurements which allows for implementing the timers away from the sensor that takes the measurements thus leading to more flexible architectures.

\begin{remark}
At each time instant $\kappa$, the optimal resource allocation problem for minimizing \eqref{eq:cost} over a finite horizon can be formulated as 
\begin{subequations} \label{eq:MIOCP}
	\begin{align}
		& \underset{\Delta_\kappa, \ldots, \Delta_K}{\min}
		& & \sum_{k=\kappa}^{K}\sum_{i\in\mathcal{N}}\tr\left(\Gamma_{i,k} {P}_{i,k|k}\right),\; \\
		&\mathrm{subject~to}
		& & {P}_{i,k|k} = \sum_{j\in\mathcal{M}}{\delta_{i,j,k}q_{i,j}}\overbar{P}_i \notag\\
		& & & \quad +(1-\sum_{j\in\mathcal{M}}{\delta_{i,j,k}q_{i,j}}) h_i(P_{i,{k-1|k-1}}),\\
		&&& i{\in}\mathcal{N}, \; k{\in}\{\kappa,\ldots,K\}, \; K{\geq}\kappa,\;\eqref{eq:constraint1},\; \eqref{eq:constraint2}.
	\end{align}
\end{subequations}
This is a mixed-integer optimal control problem (MIOCP) formulated in discrete time. \RW{Due to the extreme difficulty of solving this problem, approximate solutions can be obtained by adopting the \emph{partial outer convexification} approach and utilizing numerical solvers as discussed in \cite{Zanon:2018}}. Although the formulation in \eqref{eq:maximization} provides the solution over a single time step, in addition to computational efficiency, it facilitates distributed implementation as aforementioned.
\end{remark}

\begin{remark}
	Application of TBCoIL and its proposed variant is not limited to networks consisting of LTI processes with LQG controllers. Nevertheless, for the system described in Section~\ref{sec:model}, CoIL is only a function of locally available information and thus it enables control-aware distributed channel access with timers. The same method can be applied in other scenarios, e.g., nonlinear systems, as long as the control objective is defined such that computation of CoIL requires no explicit information exchange between subsystems.
		 %of each subsystem is proved to be calculable without requiring additional information from other subsystems. Hence, CoIL in this setting satisfies the requirements of TBCoIL, i.e., the utilized cost in the timers being a function of locally available information only, and thus it can be utilized as in \eqref{eq:timer} for control-aware distributed channel access. The same result can be obtained for nonlinear systems as long as the objective is defined such that computation of CoIL at each subsystem requires no information exchange.}
\end{remark}

\subsection{Stability Analysis}
We investigate the stability of the WNCS under the proposed channel access scheme by considering the Lyapunov mean square stability criterion. For ease of exposition, the subscript corresponding to the index of the subsystem is dropped in Definition~\ref{def:LMSS} and Lemma~\ref{lemma:bcov} since only a single subsystem is considered.
\begin{definition}[Lyapunov mean square stability \cite{Kozin:1969}] \label{def:LMSS}
The equilibrium solution is said to possess stability of the second moment if given $\varepsilon>0$, there exists $\xi(\varepsilon)$ such that $\norm{x_0}<\xi$ implies
\begin{align} \label{eq:LMSS}
	\E\{\norm{x_{k}}^2\}<\varepsilon.
\end{align}
\end{definition}

\begin{lemma} \label{lemma:bcov}
For the closed-loop systems considered in this work, there exists $\varphi$ satisfying $0< \varphi<\varepsilon$, such that \eqref{eq:LMSS} is equivalent to
\begin{align} \label{eq:trPsblt}
	\tr\left(\E\{P_{k|k}\}\right)<\varphi.
\end{align}
\end{lemma}

\begin{proof}
Let $A_L= A+BL_\infty$. The state dynamics can be written as
\begin{align*}
	x_{k+1}=A_L\hat{x}_{k|k}+Ae_{k|k}+w_{k},
\end{align*}
which yields
\begin{align*}
	\E\{\norm{x_{k+1}}^2|\mathcal{I}_{k}\}=&\tr\left(A_L\T A_L\hat{x}_{k|k}\hat{x}_{k|k}\T\right)+ \tr\left(A\T AP_{k|k}\right)\notag\\
	&+ \tr(W),
\end{align*}
since $w_k$ is zero-mean and independent of $\hat{x}_{k|k}$ and $e_{k|k}$, and $\E\{e_{k|k}|\mathcal{I}_k\}=\E\{x_{k} |\mathcal{I}_{k}\}-\hat{x}_{k|k}=0$. From the law of total expectation it follows that
\begin{align} \label{eq:bounedness}
	\E\{\norm{x_{k+1}}^2\} &= \tr\left(A_L\T A_L\E\{\hat{x}_{k|k}\hat{x}_{k|k}\T \}\right) + \tr\left(A\T A\E\{P_{k|k}\}\right) \nonumber \\
	& \quad + \tr(W),
\end{align}
which, in accordance with Definition~\ref{def:LMSS}, must be bounded. Then, for the second term we obtain \cite[Fact 8.12.28]{Bernstein:2009} 
\begin{align*}
	\tr\left(A\T A\E\{P_{k|k}\}\right) \leq \sigma_{\max}(A\T A)\tr\left(\E\{P_{k|k}\}\right),
\end{align*}
which is bounded if $\E\{P_{k|k}\}<\infty$. The certainty equivalence principle holds and thus the adopted controller ensures mean square boundedness of the state estimate, which in turn ensures boundedness of the first term in \eqref{eq:bounedness}. Since the first and last term of \eqref{eq:bounedness} are non-negative and bounded, the stability condition \eqref{eq:LMSS} only depends on the boundedness of $\E\{P_{k|k}\}$. Hence, $x_k$ is Lyapunov stable in the mean square sense if and only if there exists $0< \varphi<\varepsilon$ such that $\tr\left(\E\{P_{k|k}\}\right)<\varphi$.
%In addition, it has been proven that boundedness of $\E\{P_{k|k}\}$ guarantees that the feedback gain $L_\infty$ is stabilizing due to the perfect communication link between the controller and actuators \cite{Schenato:2007}.
\end{proof}

As a result of Lemma~\ref{lemma:bcov}, stability of the WNCS under the proposed channel access scheme can be guaranteed as long as for all $i\in\mathcal{N}$, there exists $0<\varphi_i<\infty$ such that $\tr\left(\E\{P_{i,k|k}\}\right)<\varphi_i$. To this end, we exploit the fact that the number of consecutive packet dropouts determines the error covariance at the estimator, i.e.,
\begin{align} \label{eq:PMesquita}
P_{i,k|k}=h^{t_{i,k}}(\overbar{P_i})=\sum_{c=0}^{t_{i,k}} A_i^c \overbar{P_i} {A_i\T}^c+\sum_{c=1}^{t_{i,k}} A_i^c W_i {A_i\T}^c,
\end{align}
where $\sum_{c=1}^{0} \triangleq 0$. By showing that the process $t_{i,k}$ is an ergodic Markov chain, its stationary distribution can be utilized to determine the boundedness of $\E\{P_{i,k|k}\}$. We first demonstrate how the Markov chain can be constructed and analyzed, through an illustrative example, and subsequently derive the stability conditions.

\begin{example}	\label{ex:stblty}
Consider a WNCS consisting of two unstable subsystems that share a single channel, i.e., $N=2$ and $M=1$, where the timers as set according to \eqref{eq:timer} for providing channel access. Let $\mathcal{S}=\mathbb{Z}_{\geq 0}\times \mathbb{Z}_{\geq 0}$ denote the state-space of a two-dimensional Markov chain. For any $m\in\mathbb{Z}_{\geq 0}$ and $l\in\mathbb{Z}_{\geq 0}$ we denote the respective state by $(m,l)\in\mathcal{S}$, which corresponds to $t_{1,k}=m$ and $t_{2,k}=l$. To determine the transition probabilities, we define the state-dependent action by
\begin{align} \label{eq:action}
a_{(m,l)} = \begin{cases}
0, & \text{if Subsystem } 1 \text{ claims the channel},\\
1, & \text{if Subsystem } 2 \text{ claims the channel},
\end{cases}
\end{align}
which indicates the outcome of the employing the timers in \eqref{eq:timer}. For each state $(m,l)$, CoIL for each subsystem can be determined from \eqref{eq:coil}. Furthermore, the probability of successful transmission over each wireless link is known and time-invariant. Hence, the timer values and the resulting channel access decision at each state can be determined regardless of the time instant $k$ which is represented by the state-dependent deterministic action in \eqref{eq:action}.
 
Let $0<q_i\leq 1$ be the probability of successful transmission for subsystem $i\in\{1,2\}$ and also, let $p_i\triangleq1-q_i$. Note that subscript $j$ is dropped since only a single channel is available. The transition probabilities are given by
\begin{subequations}\label{eq:PrMarkov}
\begin{align}
	&\PR\left\{ (0,l+1) \left\vert \right.  (m,l),a_{(m,l)} \right\}  \triangleq \rho_1=(1-a_{(m,l)})q_1,\label{eq:Pr1}\\
	&\PR\left\{ (m+1,0) \left\vert \right.  (m,l),a_{(m,l)} \right\} \triangleq \rho_2=a_{(m,l)} q_2,\label{eq:Pr2}\\
	&\PR\left\{ (m+1,l+1) \left\vert \right.  (m,l),a_{(m,l)} \right\} \triangleq \rho_3 \label{eq:Pr3}\\
	& \qquad \qquad =  (1-a_{(m,l)})p_1 +a_{(m,l)} p_2, \nonumber
	%\begin{aligned}[t]
	%	\rho_3=&(1-a_{(m,l)})p_1\\
	% &+a_{(m,l)} p_2, 
%	\end{aligned}\label{eq:Pr3}
\end{align}
\end{subequations}
where $(m,l)\in\mathcal{S}$ and $\PR\left\{ \cdot \left\vert \right.  \cdot,a \right\}=0$ for other cases. Fig.~\ref{fig:Markov} depicts the Markov chain modeling evolution of $t_{i,k}$'s.

\begin{figure}[t]
\centering
\includegraphics[width=.95\columnwidth]{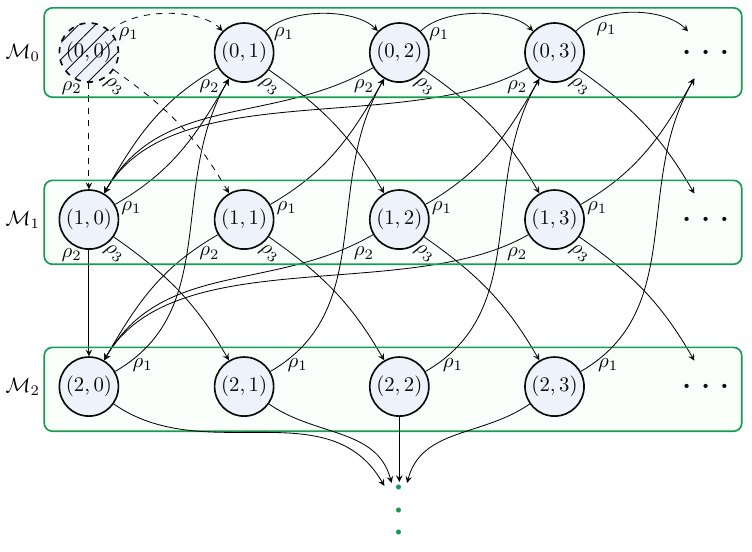}\vspace{-0.25cm}
\caption{Two-dimensional Markov chain modeling the evolution of $(t_{1,k},t_{2,k})$ in a WNCS where two subsystems share a single channel. Here, the transition probabilities $\rho_1$, $\rho_2$, and $\rho_3$ correspond to \eqref{eq:Pr1}, \eqref{eq:Pr2}, and \eqref{eq:Pr3}, respectively.\vspace{-0.25cm}} \label{fig:Markov}
\end{figure}

In order to form the transition probability matrix in a compact form, we define the \emph{state set} $\mathcal{M}_m \triangleq \cup_{{l}\in\mathbb{Z}_{\geq 0}}(m,l)$ for a given $m\in\mathbb{Z}_{\geq 0}$. In other words, $\mathcal{M}_m$ features all the states with $t_{1,k}=m$. The transition probability matrix from ${\mathcal{M}_m}$ to ${\mathcal{M}_{m+1}}$ can be expressed as
\begin{align*}
P_{{m}}=\begin{pmatrix}
	\rho_2 & \rho_3 & 0& \ldots \\
	\rho_2 & 0 & \rho_3 & \ldots \\
	\vdots & \vdots & \vdots & \ddots
\end{pmatrix}
\end{align*} 
where $P_{{m}}(l,l')=\PR\{(m+1,l')|(m,l)\}$ and first row and column indices are zero, i.e., $l,l'\geq 0$. Moreover, the transition probability matrix from ${\mathcal{M}_m}$ to ${\mathcal{M}_0}$ is given by
\begin{align*}
Q_{m}=\begin{pmatrix}
	0 & \rho_1 & 0& \ldots \\
	0 & 0 & \rho_1& \ldots \\
	\vdots& \vdots & \vdots & \ddots
\end{pmatrix}
\end{align*} 
where $Q_{{m}}(l,l')=\PR\{(0,l')|(m,l)\}$ and the row and column indexing starts from zero. As a result, the transition probability of the Markov chain can be expressed as
\begin{align*}
T=\begin{pmatrix}
	Q_0 & P_0 & \bm{0} & \ldots \\
	Q_1 & \bm{0} & P_1 & \ldots \\
	\vdots& \vdots & \vdots & \ddots
\end{pmatrix}
\end{align*}
Note that since $M<N$, the state $(0,0)$ exists only when the system is initiated and can safely be ignored. This state can be excluded from the analysis by removing the first row and column of $T$, which is denoted by $\hat{T}$. Thus, the resulting Markov chain has a single communicating class and is irreducible, aperiodic and, since $q_i>0$, it is positive recurrent. As a result \cite[Ch.~1]{Norris:1997}, this chain always has a limiting distribution $\bm\pi=[\bm{\pi_0},\bm{\pi_1},\ldots]$ where $\bm{\pi_m}=[\pi_{(m,0)},\pi_{(m,1)},\ldots]$, which is the unique solution to 
\begin{align} \label{eq:pi}
\bm{\pi}\hat{T}=\bm{\pi} \qquad \bm{\pi}\bm{1}=1,
\end{align}
where $\bm{1}$ is the all-ones column vector of appropriate dimensions. The vector $\bm\pi$ found by solving \eqref{eq:pi} can be used to determine $\mu_i(t)\triangleq \PR\{t_{i,k}=t\}$ which is crucial in the remaining of this section. For the illustrative case considered in this example we have
\begin{align}
\mu_1(t)&\triangleq \PR\{t_{1,k}=t\}= \sum_{l=0}^{\infty}\pi_{(t,l)}, \label{eq:mu1}\\
\mu_2(t)&\triangleq \PR\{t_{2,k}=t\}= \sum_{m=0}^{\infty}\pi_{(m,t)}, \label{eq:mu2}
\end{align}
\end{example}
The method of Example~\ref{ex:stblty} can readily be applied in larger WNCSs to form the Markov chain that models the evolution of $t_{i,k}$'s. The states of the chain in such general settings represent $(m_1,m_2,\ldots,m_N)$, where $m_i=t_{i,k}$. Furthermore, the transition probabilities are determined by the state-dependent actions that result from the interaction of $N{\times}M$ state-dependent timers. This leads to an $N$-dimensional irreducible, aperiodic and positive recurrent Markov chain with a corresponding transition probability matrix $\hat{T}$. Therefore, the limiting distribution of this Markov chain can be used to determine the unique solution of \eqref{eq:pi}. Therefore, ${\mu}_i(t)$ can be determined for all $i$ and $t$, which, as we show next, is crucial for examining whether $\tr\left(\E\{P_{k|k}\}\right)$ is bounded as required by Lemma~\ref{lemma:bcov}.

%\TF{The method of Example~\ref{ex:stblty} can readily be applied to larger WNCSs to form the Markov chain that models the evolution of $t_{i,k}$'s. In such settings, each state can be described by $(m_1,m_2,\ldots,m_N)$ where $m_i=t_{i,k}$, and the corresponding $NM$ state-dependent timers can be used to determine the transition probabilities. This leads to an $N$-dimensional irreducible, aperiodic and positive recurrent Markov chain with the transition probability matrix $\hat{T}$. Therefore, the unique solution of \eqref{eq:pi} can be obtained which yields the limiting distribution of this Markov chain and thus ${\mu}_i(t)$ for all $i$ and $t$. As it will be shown next, this is a crucial step for checking whether $\tr\left(\E\{P_{k|k}\}\right)$ is bounded as required by Lemma~\ref{lemma:bcov}.}

\begin{theorem} \label{theorem:1}
The proposed channel access method stabilizes the WNCS in the sense of Definition~\ref{def:LMSS}, if for all $i\in\mathcal{N}$, the following condition holds
	\begin{align} \label{eq:stblty}
		\lim\limits_{t\to\infty} \mu_i(t)^{1/t} <\frac{1}{{\sigma_{\max}^2(A_i)}}.
	\end{align}
\end{theorem}

\begin{proof}
Due to the ergodicity of the Markov chain, taking the limit of the expected value of \eqref{eq:PMesquita} yields
	\begin{align} \label{eq:conv1}
		\lim\limits_{k\to\infty} \E\{P_{i,k|k}\} =& \sum_{t=0}^\infty \mu_i(t)\sum_{c=0}^t( A_i^c \overbar{P_i} {A_i\T}^c) \notag \\
		&+ \sum_{t=0}^\infty \mu_i(t)\sum_{c=1}^t( A_i^c W_i {A_i\T}^c).
	\end{align}
Subsequently,
\begin{align} \label{eq:normEP}
	\norm{\lim\limits_{k\to\infty} \E\{P_{i,k|k}\}}\leq (\norm{\overbar{P_i}}+\norm{W_i}) \sum_{t=0}^\infty \mu_i(t)\sum_{c=0}^{t} \norm{A_i^c}^2.
\end{align}
Similar to the proof of \cite[Theorem 1]{Mesquita:2012}, by Cauchy's root test, this series is convergent if
\begin{align} \label{eq:Cauchy}
	\lim\limits_{t\to\infty} \mu_i(t)^{1/t}\norm{A_i^t}^{2/t}<1,
\end{align}
and applying Gelfand's formula yields
\begin{align}
	\sigma_{\max}^2(A_i)\lim\limits_{t\to\infty} \mu_i(t)^{1/t}<1.
\end{align}
Hence, if \eqref{eq:stblty} holds for all $i\in\mathcal{N}$, the upper bound in \eqref{eq:normEP} exists which itself guarantees that $0<\varphi_i<\infty$ exists such that $\tr\left(\E\{P_{i,k|k}\}\right)<\varphi_i$ thus concluding the proof.
\end{proof}
In case the closed form expression for $\mu_i(t)$ is known, Theorem~\ref{theorem:1} can readily be utilized to verify stability. In general, however, finding a closed form expression might not be possible as it is the case in Example~\ref{ex:stblty}. Nevertheless, as it will be demonstrated in Section~\ref{sec:numericals}, the \emph{p-series convergence test} can be used in practice to examine stability within the same framework.

\begin{remark}
It should be noted that although one of the transition probabilities in \eqref{eq:PrMarkov} is inevitably zero, all states can be reached with a nonzero probability. This is due to the fact that $h_i^t(X)$ is a monotonically increasing function of $t$ \cite[Lemma  A.3]{Shi:2010}. More specifically, assume that for a given state $(m,l)$, the parameters are such that Subsystem $1$ has a smaller timer which means that $a=0$. From \eqref{eq:PrMarkov} it follows that $\PR\left\{ (m+1,0) \left\vert \right.  (m,l),a \right\} =0$. Nonetheless, there exists a state $(m,l')$ with $l'>l$ such that CoIL of Subsystem $2$ is large enough to result in a smaller timer value than Subsystem $1$. Therefore, in state $(m,l')$ the actions is $a=1$ and thus $\PR\left\{ (m+1,0) \left\vert\right. (m,l'),a \right\} =q_2$.
\end{remark}

% =====================================================
%
%
% Scheduling with learning the channel parameters
%
%
% =====================================================
\section{Channel access over unknown memoryless channels}\label{sec:MAB}

\subsection{Problem Statement}
Optimal resource allocation requires knowledge of the exact values of $q_{i,j}$'s which describe the time-invariant distributions of the time-varying channels. Due to the dynamic nature of the considered subsystems and the changing environment, the coherence time of the channel is relatively small and fast fading occurs, rendering it impossible to have instantaneous channel state information (CSI) acquisition. In such settings, learning methods can be applied to gain knowledge of the underlying channel statistical parameters, which are assumed to change very slowly with respect to the coherence time. Despite the abundance of existing learning algorithms which are applicable to standard wireless networks, adopting a suitable learning algorithm in our problem is challenging due to two setup-related reasons: \emph{(i)} the considered WNCS structure allows no information exchange between subsystems and thus the learning method should be compatible with distributed implementation; \emph{(ii)} since the main objective is minimizing the quadratic cost, the adopted algorithm should be compatible with the proposed timer-based mechanism. More specifically, channel statistics cannot be learned separately without taking into account CoIL. We aim at devising a novel distributed method which aims at maintaining a good control performance while learning the channel statistics.

\subsection{A MAB Approach}
MAB problem refers to optimal sequential allocation in unknown random environments. In classic single-player stochastic MAB, a player has access to multiple, say $M$, independent arms. The player pulls an arm $j\in\mathcal{M}$ at each round which yields a reward drawn randomly from an unknown probability distribution specific to that arm. Since the player has no prior knowledge of the reward distributions, he might play an inferior arm in terms of reward. We define \emph{regret} as the difference between the reward obtained from playing the best arm and the player's choice. Let $r_{j,k}$ and $\mathrm{I}_k$ denote the instantaneous reward obtained from arm $j$ and the selected arm at round $k$, respectively. Then, the (external) regret up to round $K$ is defined by
\begin{align}
R_K =\max_{j\in\mathcal{M}} \E\left\{\sum_{k=1}^{K}{r_{j,k}}\right\} - \E\left\{\sum_{k=1}^{K}{{r_{\mathrm{I}_k}}}\right\}.
\end{align}
The objective is to find a \emph{policy} for selecting the arms, i.e., to determine $\mathrm{I}_k$ at each round $k$, such that this regret is minimized over the game horizon. The performance of a policy relies on how it addresses the \emph{exploration/exploitation} dilemma: searching for a balance between exploring all arms to learn their reward distribution while playing the best arm more often to gain more reward.

The channel selection problem for a single subsystem can be conveniently cast as a single-player MAB. In this scenario, channels represent arms and playing an arm corresponds to claiming a channel for packet transmission. We adopt a binary rewarding scheme ($r_{j,k}\in\{0,1\}$), where in case of a successful transmission, a unit reward is obtained over the corresponding channel {($r_{\mathrm{I}_k}=1$)}, otherwise, no reward is earned ({$r_{\mathrm{I}_k}=0$}). The channels are independent and packet dropouts are i.i.d. random and, subsequently, the rewards are i.i.d. random. The mean of the Bernoulli distribution of rewards over each channel corresponds to the probability of successful transmission \eqref{eq:Psucc}. Therefore, by adopting a suitable policy, after an initial exploration phase, the channel with the best quality is exploited for maximizing the success rate or, equivalently, the reward.

\emph{Index policies} are a class of solutions to this problem, which assign an index to each arm and play the one with the largest index. One of the main categories of the methods that belong to this class are based on \emph{upper confidence bound} (UCB). These policies estimate an upper bound of the mean reward of each arm at some fixed confidence level and determine the indices accordingly. One of the celebrated results based on this idea is UCB1, a policy introduced in \cite{Auer:2002}. In this policy, at each round $k$, the upper confidence bound of the mean reward, denoted by $\hat{q}_{j,k}$, is calculated and the arm with the largest $\hat{q}_{j,k}$ is played. In this work, we use a slightly modified version of UCB1 to ensure collision-free channel access. More specifically, we calculate $\hat{q}_{j,k}$ by
\begin{align} \label{eq:ucb1}
\hat{q}_{j,k} = \bar{r}_{j,k}+\sqrt{\frac{2\ln z_k}{z_{j,k}+\epsilon_{j,k}}},
\end{align}
\RW{where, similar to the UCB1 algorithm, $z_k$ is the total number of plays, $z_{j,k}$ denotes the number of plays of arm $j$ up to $k$, and $\bar{r}_{j,k}$ is the average reward obtained from playing arm $j$ up to $k$, i.e.,}
\begin{align*}
	\bar{r}_{j,k} = \frac{\sum_{\kappa=1}^{k} r_{j,\kappa} {\mathbbm{1}}_{{\mathrm{I}_\kappa=j}} }{z_{j,k}},
\end{align*}
\RW{where {$\mathbbm{1}_{\mathrm{I}_\kappa=j}$} is $1$ when arm $j$ is played at round $\kappa$ is $j$. Moreover, differently from the original algorithm, we also employ a uniformly distributed random variable $\epsilon_{j,k}\sim\mathcal{U}(a,b)$ in the exploration term of \eqref{eq:ucb1}. This ensures that the upper confidence bounds are distinct during the initial exploration phase of the algorithm thus enabling collision-free channel access with timers. Furthermore, since $z_{j,k}\in\mathbb{Z}_{\geq 0}$ choosing small values for $a$ and $b$ ensures that convergence of \eqref{eq:ucb1} is not disrupted as demonstrated in Subsection~\ref{subsection:res-MAB}. Note that the index of the played arm at $k$ is given by} 
\begin{align*}
{\mathrm{I}_k} = \argmax_{j\in\mathcal{M}} \hat{q}_{j,k}.
\end{align*}
%The proposed expression for calculating the upper confidence bound enables achieving similar performance to UCB1 while the random variable $\epsilon_{j,k}\sim\mathcal{U}(a,b)$ ensures that obtaining identical upper confidence bounds at each round has Lebesgue measure zero. This property will be exploited for enabling collision-free distributed implementation. Furthermore, since $z_{j,k}\in\mathbb{Z}_{\geq 0}$ choosing $-1<a,b<1$ ensures that convergence of \eqref{eq:ucb1} is not disrupted.

The problem of distributed channel access in standard wireless networks, unlike WNCSs, only concerns maximizing throughput without considering the importance of the contents of the data packets \cite{Maghsudi:2016}. This problem can be cast as a multi-player MAB where, given the aforementioned binary rewarding scheme is adopted, the maximum reward at each time step $k$ is given by optimal resource allocation according to
\begin{align} \label{eq:bestSubCh}
\max_{\delta_{i,j,k}\in\{0,1\}}\sum_{i\in\mathcal{N}}\sum_{j\in\mathcal{M}}q_{i,j}\delta_{i,j,k},
\end{align}
subject to constraints \eqref{eq:constraint1} and \eqref{eq:constraint2}. Since the reward distribution over each wireless link is assumed to be time-invariant, the optimal decision variables are likewise time-invariant. Consequently, subscript $k$ is dropped and we denote the solution by $\delta_{i,j}^{q*}$. As a result, regret is given by
\begin{align} \label{eq:regret1}
R_K = K\sum_{i\in\mathcal{N}}\sum_{j\in\mathcal{M}}q_{i,j}\delta_{i,j}^{q*} - \sum_{k=1}^{K}\sum_{i\in\mathcal{N}} q_{i,\mathrm{I}_{i,k}},
\end{align}
where $\mathrm{I}_{i,k}$ denotes the index of the selected channel by subsystem $i$ at round $k$. By implementing suitable policies one can ensure that this regret grows logarithmically.

\subsection{Distributed Channel Access Algorithm}
We first cast our problem as a multi-player MAB and then propose a novel indexing policy for addressing the exploration/exploitation dilemma with respect to the control performance in a distributed manner. Since our goal is to minimize the quadratic cost, with a slight abuse of notation, we define the \emph{cost regret} up to time $K$ as 
\begin{align} \label{eq:regret2}
R_{\mathop{cost},K} =& \sum_{k=1}^{K}\min_{\mathcal{F}_k}\E\{{J}_k|\mathcal{I}^{k-1},\mathcal{F}_k\} \nonumber\\
	&- \sum_{k=1}^{K}\min_{\mathcal{F}_k}\E\{{J}_k|\mathcal{I}^{k-1},\mathcal{F}_k,\mathcal{Q}\}.
\end{align}
The aim of the policy is to, without any prior knowledge of the channel qualities, determine the subset of subsystems that transmit and their respective channels; this corresponds to the first term of \eqref{eq:regret2}. The last term of \eqref{eq:regret2} is the minimum cost that is incurred when $\mathcal{Q}$ is known; its solution is obtained by solving the optimal resource allocation problem formulated in \eqref{eq:maximization}. Performance of a channel access policy can now be measured in terms of minimizing the cost regret. 

Although the cost regret is fundamentally different from the standard regret defined in \eqref{eq:regret1}, we propose a new method for exploiting the well-established results for minimizing the latter in our favor by introducing time-varying weights that reflect the control performance. We still apply the aforementioned binary rewarding scheme, i.e.,
\begin{align} \label{eq:costregrew}
r_{i,j,k} = \begin{cases}
	1, & \text{if transmission is successful,}\\%\delta_{i,j,k}=1 \text{ and }\gamma_{i,j,k}=1, \\
	0, & \text{otherwise},
\end{cases}
\end{align}
which is an i.i.d. random variable with $\E\{r_{i,j,k}\} = q_{i,j}$, and calculate the initial index of each channel by an index policy that is compatible with distributed implementation. More specifically, policies which require local information for calculating the index of each arm and the resulting indices are distinct, e.g., as per \eqref{eq:ucb1}. Nevertheless, in our policy, these initial indices are then weighted by the control performance metric, namely CoIL. Consequently, the index of the selected channel by each subsystem is given by
\begin{align} \label{eq:tauq}
{\mathrm{I}_{i,k}} = \begin{cases}
	\displaystyle\argmax_{j\in\mathcal{M}} \delta_{i,j,k}^* , & \text{if~} \exists j: \delta_{i,j,k}^* \neq 0,\\
	\emptyset, & \text{otherwise},
\end{cases}
\end{align}
where $\delta_{i,j,k}^*$ is obtained by the following optimization problem
\begin{align} \label{eq:maximization2_themis}
\bm{\delta}_k^* & \triangleq \begin{bmatrix}\delta_{i_1,1,k}^* & \ldots& \delta_{i_M,M,k}^* \end{bmatrix}\T  \nonumber \\ 	
&=\argmax_{\delta_{i,j,k}\in\{0,1\}}\sum_{i\in\mathcal{N}}\sum_{j\in\mathcal{M}}{\CoIL}_{i,k}\hat{q}_{i,j,k}\delta_{i,j,k},
\end{align}
subject to constraints \eqref{eq:constraint1} and \eqref{eq:constraint2}. This ensures correct estimation of the success probability of each channel, while at the same time, the slot is allotted to the subsystem with the highest cost. This policy can be implemented in a distributed manner by adopting the timer-based mechanism for solving \eqref{eq:maximization2_themis}. By using the weighted indices as the local cost, the timers are determined by % \HW{Note that the distribution of the reward in \eqref{eq:costregrew} is time-invariant as $q_{i,j}$'s are time-invariant.}
\begin{align}\label{eq:timer-ucb}
\tau_{i,j,k} = \frac{\lambda}{\CoIL_{i,k}\hat{q}_{i,j,k}}.
\end{align}
As a result, assuming that the duration of the flag packet is negligible, since $\hat{q}_{i,j,k}$ has Lebesgue measure zero, this mechanism guarantees collision-free channel access even for homogeneous systems.

When the channel access policy is designed with respect to regret as defined in \eqref{eq:regret1}, its implementation only maximizes the number of successful transmission. This translates to sacrificing the control performance, which is the primary objective in WNCSs, in favor of maximizing the throughput. Nevertheless, the outcome of these policies can be manipulated in favor of the control objective by applying the time-varying weights, i.e., CoIL. This significantly improves performance despite (possibly) higher packet dropout rates as shown by the numerical results in Section \ref{sec:numericals}. Algorithm~\ref{alg:coil-ucb} illustrates the detailed distributed implementation of our proposed policy.

\begin{remark}
	Before initiating the index calculation in \eqref{eq:ucb1}, each arm needs to be played once. This can easily be achieved by temporarily adopting round-robin, where subsystems transmit according to a random sequence for the first $N{\times}M$ time steps, i.e., the number of subsystems times the number of channels. Afterwards, by setting the reference time to $N{\times}M$, the generated set of observations and accumulated rewards, denoted by $\mathcal{Z}_{i,1} \triangleq  \left\{z_{i,j,1}|\forall j\in\mathcal{M}\right\}$ and $\mathcal{R}_{i}\triangleq \left\{R_{i,j}|\forall j\in\mathcal{M}\right\}$, respectively, are used for determining channel access according to Algorithm~\ref{alg:coil-ucb}.
\end{remark}

\begin{algorithm} \label{alg:coil-ucb}
\SetAlgoLined
\DontPrintSemicolon
\KwIn{channel indices $\mathcal{M}$, constant value for timer setup $\lambda$, the initial observation history $\mathcal{Z}_{i,1}$ and accumulated rewards $\mathcal{R}_i$.}
\For{$k=1, 2, \ldots$}{
	$z_{i,k} = \sum_{j\in\mathcal{M}}z_{i,j,k}$ and $\bar{r}_{i,j,k} = \frac{R_{i,j}}{z_{i,j,k}},\forall j\in\mathcal{M}$ \;	
	randomly generate $\epsilon_{j,k}\in[-0.5,0.5],\forall j\in\mathcal{M}$\;		
	calculate $\hat{q}_{i,j,k}$ \eqref{eq:ucb1} and	$\CoIL_{i,k}$ \eqref{eq:coil}\; 
	start $\mathrm{timer}_{i,j,k}$ from $\tau_{i,j,k}$ \eqref{eq:timer-ucb} \;
	initiate set of dummy indices $\mathcal{F}_i =  \{1,\ldots,M\}$\;
	\While{$\mathcal{F}_i\neq \emptyset$}{
		\For{$j \leftarrow 1$ \KwTo $M$}{
			\uIf{$\mathrm{timer}_{i,j,k} \neq 0$ and still running}{% 
				listen for signals\;
				\If{signal is received in channel $j$}{
					stop $\mathrm{timer}_{i,j,k}$ and $\mathcal{F}_i\leftarrow\mathcal{F}_i\backslash\{j\}$\;
				}
			}
			\ElseIf{$\mathrm{timer}_{i,j,k}=0$}{
				send flag on channel $j$\\
				$\mathcal{F}_{i}\leftarrow \emptyset$ and ${\mathrm{I}_{i,k}} \leftarrow j$\\
				freeze all running timers\;
			}		
		}
	}
	$\mathcal{Z}_{i,k+1} \leftarrow \mathcal{Z}_{i,k}$ \;
	\If{${\mathrm{I}_{i,k}}\neq\emptyset$
	}{
		transmit on channel ${\mathrm{I}_{i,k}}$\;
		$z_{i,{{\mathrm{I}}_{i,k}},k+1} \leftarrow z_{i,{{\mathrm{I}}_{i,k}},k+1}+1$\;
		\lIf{$\gamma_{i,{{\mathrm{I}}_{i,k}},k} = 1$}{$R_{i,{{\mathrm{I}}_{i,k}}} \leftarrow R_{i,{{\mathrm{I}}_{i,k}}}+1$}
	}
}
\caption{Implementation of the timer-based channel access mechanism at subsystem $i$}
\end{algorithm}
%\vspace{-.2cm}

% =====================================================
%
%
% NUMERICAL RESULTS
%
%
% =====================================================
\section{Numerical Results}\label{sec:numericals}
The results presented in this section are obtained by considering homogeneous WNCSs consisting of identical two-wheeled balancing robots. The matrices in \eqref{eq:process} are obtained from the continuous-time model presented in \cite{Pezzutto:2020} with a sampling of $0.02$ seconds which yields
\begin{align*}
	A{=}\setlength\arraycolsep{2.5pt}\begin{bmatrix}
		1  & 0.009 &  0.019 &  0.001\\
		0  & 1.011 &  0.000 &  0.020\\
		0  & 0.879 &  0.928 &  0.073\\
		0  & 1.101 &  0.037 &  0.968
	\end{bmatrix},
	B{=}\begin{bmatrix}
		0.001\\
		-0.001\\
		0.093\\
		-0.062
	\end{bmatrix},
	C{=}\setlength\arraycolsep{2.5pt}\begin{bmatrix}
		1 & 0 \\
		0 & 1 \\
		0 & 0\\
		0 & 0 \\
	\end{bmatrix}^T.
\end{align*}
%\begin{align*}
%		A&=\begin{bmatrix}
%			1  & 0.0088 &  0.0193 &  0.0007\\
%			0  & 1.0110 &  0.0004 &  0.0196\\
%			0  & 0.8788 &  0.9280 &  0.0729\\
%			0  & 1.1009 &  0.0372 &  0.9681
%		\end{bmatrix}, \quad
%	B=\begin{bmatrix}
%		0.0009\\
%		-0.0006\\
%		0.0925\\
%		-0.0620
%	\end{bmatrix}, \\
%C&=\begin{bmatrix}
%	1 & 0 & 0 & 0\\
%	0 & 1 & 0 & 0\\
%\end{bmatrix}.
%\end{align*}
The states are the wheel angle, the tilt angle, and their respective derivatives. The input is the voltage of the DC motors delivering torque to the wheels and the output is the measurements given by the encoder and the inertial measurement sensor. Furthermore, the covariance of the process disturbance and measurement noise are chosen as $W=0.1I_4$ and $V=0.01I_2$, respectively, and the weighting matrices in \eqref{eq:cost} are $Q=I_4$ and $R=0.1$.

\subsection{Stability Analysis}
\begin{figure}[t]
	\centering
	\includegraphics[width=0.9\columnwidth]{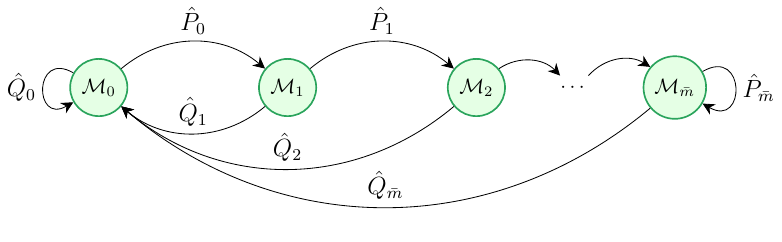}
	\vspace{-0.32cm}
	\caption{Graphical representation of the truncated version of the Markov chain depicted in Fig.~\ref{fig:Markov}.\vspace{-0.3cm}}\label{fig:Markovres}	
\end{figure}

Considering the scenario of two balancing robots contending for channel access as in Example~\ref{ex:stblty}, the truncated Markov chain can be analyzed to provide insight on the stability of the system. Due to the lack of a closed form expression for $\mu_i(t)$, the stability condition in Theorem~\ref{theorem:1} cannot be verified directly. Therefore, we consider a truncated version of the Markov chain of Fig.~\ref{fig:Markov} by letting $0\leq m,l \leq \overbar{m}$ as graphically represented in Fig.~\ref{fig:Markovres}. In this scenario, we can form the transition matrices $\hat{P}_{{m}}$ and $\hat{Q}_{{m}}$ by keeping only the first $\overbar{m}$ rows and columns of $P_{{m}}$ and $Q_{{m}}$, respectively. As a result, the transition probability matrix of this chain can be expressed as
\begin{align*}
	\hat{T}=\begin{pmatrix}
		\hat{Q}_0 & \hat{P}_0 & \bm{0} & \ldots& \bm{0} \\
		\hat{Q}_1 & \bm{0} & \hat{P}_1 & \ldots& \bm{0} \\
		\vdots& \vdots & \vdots & \ddots & \vdots\\
		\hat{Q}_{\overbar{m}} & \bm{0} &\bm{0} & \ldots & \hat{P}_{\overbar{m}}
	\end{pmatrix},
\end{align*}
which is row stochastic, irreducible, and aperiodic; therefore, the stationary probability vector can be obtained by  \cite{Krikidis:2012, Xu:2016} %, 
\begin{align} \label{eq:buffer}
	\bm{\pi}=\bm{1}\T (\hat{T}-I_{\overbar{m}}+{D})^{-1},
\end{align}
where ${D}(m,l)=1$ for all $m,l$. Although this is an approximation of the actual chain in Fig.~\ref{fig:Markov}, by choosing sufficiently large $\overbar{m}$, \eqref{eq:buffer} provides a highly accurate approximation of the stationary distribution of the actual chain. According to the p-series convergence test, the series on the right hand side of \eqref{eq:normEP} is convergent if exists $p>1$ and $\beta<\infty$ such that
\begin{align} \label{eq:convres}
 	\lim\limits_{\overbar{m}\to\infty} \sum_{t=0}^{\overbar{m}}\mu_i(t)\norm{A_i}^{2t} \leq \lim\limits_{\overbar{m}\to\infty} \sum_{t=0}^{\overbar{m}} \frac{\beta}{t^p}.
\end{align}

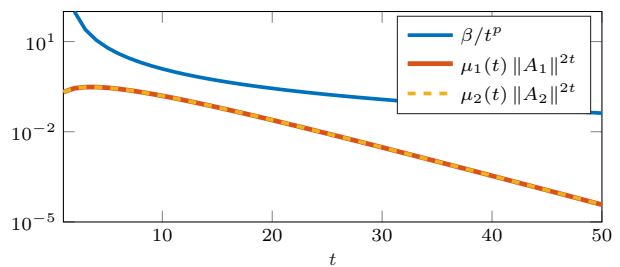
\begin{figure} [h]
	\scriptsize
	\centering 
	\setlength\fheight{2.8cm} 
	\setlength\fwidth{0.85\columnwidth}
\definecolor{mycolor1}{rgb}{0.00000,0.44700,0.74100}%
\definecolor{mycolor2}{rgb}{0.85000,0.32500,0.09800}%
\definecolor{mycolor3}{rgb}{0.92900,0.69400,0.12500}%
\begin{tikzpicture}

\begin{axis}[%
width=0.951\fwidth,
height=\fheight,
at={(0\fwidth,0\fheight)},
scale only axis,
xmin=1,
xmax=50,
xlabel style={font=\color{white!15!black}},
xlabel={$t$},
ymode=log,
ymin=1e-05,
ymax=100,
yminorticks=true,
axis background/.style={fill=white},
legend style={legend cell align=left, align=left, draw=white!15!black}
]
\addplot [color=mycolor1, line width=1.4pt]
  table[row sep=crcr]{%
2	100\\
3	25\\
4	11.1111111111111\\
5	6.25\\
6	4\\
7	2.77777777777778\\
8	2.04081632653061\\
9	1.5625\\
10	1.23456790123457\\
11	1\\
12	0.826446280991734\\
13	0.694444444444444\\
14	0.59171597633136\\
15	0.510204081632653\\
16	0.444444444444444\\
17	0.390625\\
18	0.346020761245677\\
19	0.308641975308641\\
21	0.25\\
23	0.206611570247934\\
25	0.173611111111111\\
27	0.14792899408284\\
29	0.127551020408163\\
31	0.111111111111111\\
34	0.0918273645546371\\
37	0.0771604938271603\\
40	0.0657462195923733\\
43	0.056689342403628\\
46	0.0493827160493827\\
50	0.041649312786339\\
51	0.04\\
};
\addlegendentry{${\beta}/{t^{p}}$}

\addplot [color=mycolor2, line width=1.8pt]
  table[row sep=crcr]{%
1	0.209523809521113\\
2	0.27904198476288\\
3	0.306219564528316\\
4	0.306776079101337\\
5	0.291439280592813\\
6	0.26741414332379\\
7	0.239438242607958\\
8	0.210536731645512\\
9	0.182559733082883\\
10	0.156562612962386\\
11	0.133073197881098\\
12	0.112277967462111\\
14	0.0785386186047745\\
16	0.0539568800174694\\
18	0.0365708699011607\\
20	0.0245294935812066\\
23	0.0132752214436822\\
26	0.00708848364631642\\
30	0.00302477576061917\\
34	0.0012749116442097\\
39	0.000427827945707174\\
46	9.13934774025264e-05\\
48	5.86127238732671e-05\\
49	4.68545203889843e-05\\
51	3.03043954749461e-05\\
};
\addlegendentry{$\mu_1(t)\norm{A_1}^{2t}$}

\addplot [color=mycolor3, dashed, line width=1.4pt]
  table[row sep=crcr]{%
1	0.20952380951955\\
2	0.279041984761633\\
3	0.306219564524784\\
4	0.306776079102858\\
5	0.291439280598936\\
6	0.267414143326585\\
7	0.23943824261072\\
8	0.210536731649673\\
9	0.182559733074078\\
10	0.156562612961302\\
11	0.133073197898434\\
12	0.112277967458198\\
14	0.0785386186412969\\
16	0.0539568801059081\\
18	0.036570869898899\\
20	0.0245294935791378\\
23	0.013275221794928\\
26	0.00708848306176855\\
30	0.00302477670894359\\
34	0.00127492026241582\\
39	0.000427840093156777\\
47	7.33882846622085e-05\\
51	3.0139117114372e-05\\
};
\addlegendentry{$\mu_2(t)\norm{A_2}^{2t}$}

\end{axis}
\end{tikzpicture}%
	\vspace{-0.3cm}
	\caption{Convergence analysis of the left hand side \eqref{eq:convres} by element-wise comparison with the p-series ($\beta=100$, $p=2$) given that $q_1=0.40$ and $q_2=0.44$.} \vspace{-0.25cm}
	\label{fig:stblty} 
	%	\normalsize
\end{figure}

Fig. \ref{fig:stblty} depicts the values of $\mu_i(t)\norm{A_i}^{2t}$ and ${\beta}/{t^p}$ as a function of $t$ given that $p=2$ and $\beta=100$. The wireless link qualities are assumed to be $q_1=0.40$ and $q_2=0.44$, and $\overbar{m}=52$ is chosen for computing \eqref{eq:buffer} which is then used to calculate $\mu_1(t)$ and $\mu_2(t)$ as per \eqref{eq:mu1} and \eqref{eq:mu2}, respectively. As it can be seen, $\mu_i(t)\norm{A_i}^{2t}$ is a monotonically decreasing function of $t$ for $t\geq 4$ for both subsystems and it is upperbounded by $\frac{\beta}{t^p}$. This can be construed as convergence of the series in \eqref{eq:convres} and thus stability. In sharp contrast, Fig.~\ref{fig:stblty_u} shows case where $q_1$ is reduced to $0.2$. In this scenario, $\mu_i(t)\norm{A_i}^{2t}$ becomes an increasing function of $t$ which indicates that the right hand side of \eqref{eq:normEP} is a divergent series and thus stability of the system cannot be guaranteed.

\begin{figure} [h]
	\scriptsize
	\centering 
	\setlength\fheight{2.8cm} 
	\setlength\fwidth{0.85\columnwidth}
\definecolor{mycolor1}{rgb}{0.00000,0.44700,0.74100}%
\definecolor{mycolor2}{rgb}{0.85000,0.32500,0.09800}%
\definecolor{mycolor3}{rgb}{0.92900,0.69400,0.12500}%
\begin{tikzpicture}

\begin{axis}[%
width=0.951\fwidth,
height=\fheight,
at={(0\fwidth,0\fheight)},
scale only axis,
xmin=1,
xmax=50,
xlabel style={font=\color{white!15!black}},
xlabel={$t$},
ymode=log,
ymin=0.01,
ymax=100,
yminorticks=true,
axis background/.style={fill=white},
legend style={legend cell align=left, align=left, draw=white!15!black}
]
\addplot [color=mycolor1, line width=1.4pt]
  table[row sep=crcr]{%
2	100\\
3	25\\
4	11.1111111111111\\
5	6.25\\
6	4\\
7	2.77777777777778\\
8	2.04081632653061\\
9	1.5625\\
10	1.23456790123457\\
11	1\\
12	0.826446280991734\\
13	0.694444444444444\\
14	0.59171597633136\\
15	0.510204081632653\\
16	0.444444444444444\\
17	0.390625\\
18	0.346020761245677\\
19	0.308641975308641\\
20	0.277008310249308\\
21	0.25\\
22	0.226757369614512\\
23	0.206611570247934\\
24	0.189035916824196\\
26	0.16\\
28	0.137174211248285\\
30	0.118906064209274\\
32	0.104058272632675\\
34	0.0918273645546371\\
36	0.0816326530612245\\
38	0.0730460189919649\\
40	0.0657462195923733\\
42	0.0594883997620459\\
45	0.0516528925619834\\
48	0.045269352648257\\
51	0.04\\
};
\addlegendentry{${\beta}/{t^{p}}$}

\addplot [color=mycolor2, line width=1.8pt]
  table[row sep=crcr]{%
1	0.119624674574242\\
2	0.159315099769793\\
3	0.193503109696623\\
4	0.243780603114664\\
5	0.287786036156351\\
6	0.327539606196333\\
7	0.381417258148891\\
8	0.430572681940968\\
9	0.476793403600462\\
10	0.521450984712571\\
11	0.565609273103171\\
12	0.610105300984489\\
14	0.702672488643798\\
17	0.857549297461555\\
22	1.1819109879479\\
40	3.70166111383043\\
51	7.43335590007883\\
};
\addlegendentry{$\mu_1(t)\norm{A_1}^{2t}$}

\addplot [color=mycolor3, dashed, line width=1.4pt]
  table[row sep=crcr]{%
1	0.17682571594731\\
2	0.235494948502984\\
4	0.271423950157921\\
5	0.269554345966408\\
6	0.27255221541597\\
7	0.279467658596421\\
8	0.289611042414389\\
9	0.302488072064118\\
10	0.317751447630415\\
12	0.354576433036237\\
14	0.399093409260856\\
17	0.480074948345448\\
24	0.746216252153675\\
50	3.87653288169258\\
51	2.89113031337642\\
};
\addlegendentry{$\mu_2(t)\norm{A_2}^{2t}$}

\end{axis}
\end{tikzpicture}%
	\vspace{-0.3cm}
	\caption{Convergence analysis of the left hand side \eqref{eq:convres} by element-wise comparison with the p-series ($\beta=100$, $p=2$) given that $q_1=0.20$ and $q_2=0.44$.} \vspace{-0.25cm}
	\label{fig:stblty_u} 
	%	\normalsize
\end{figure}

\subsection{MAB Approach} \label{subsection:res-MAB}
\begin{figure}[t]
\centering
\includegraphics[width=0.91\columnwidth]{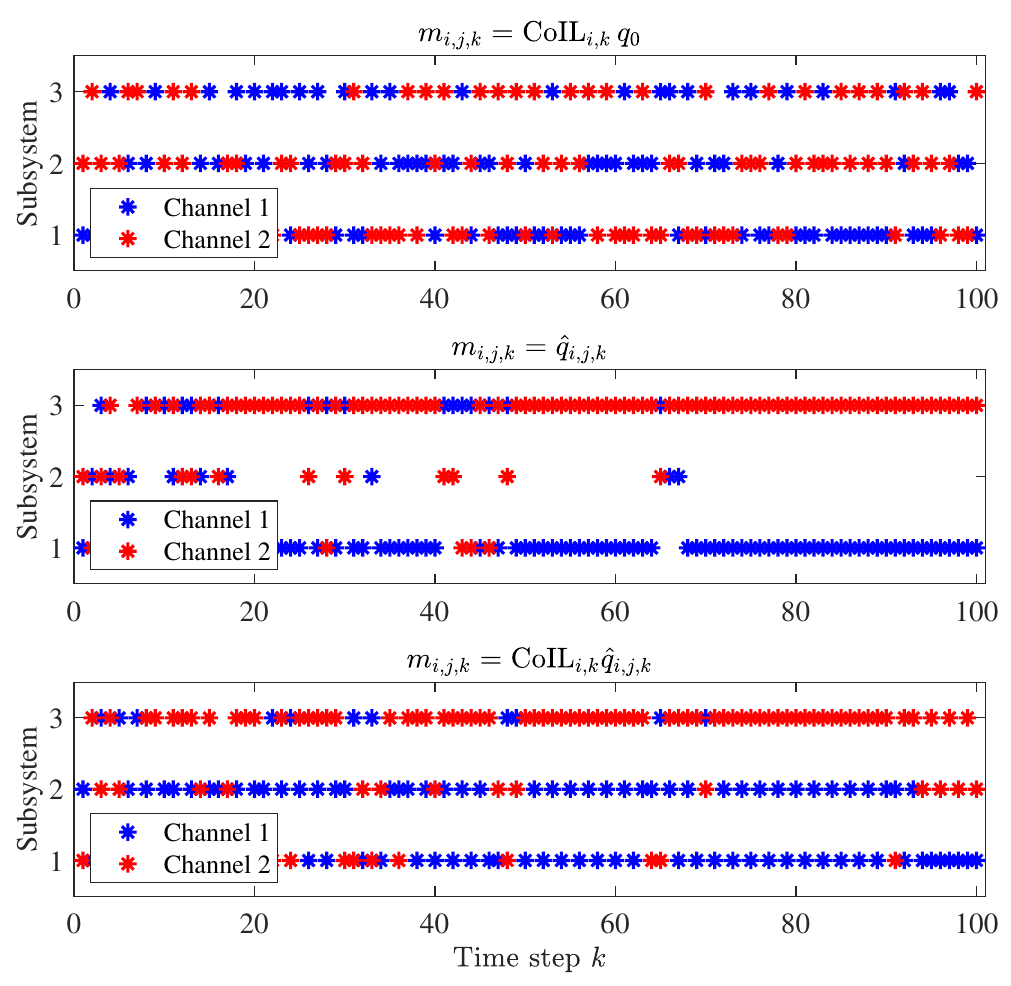}
\vspace{-0.3cm}
\caption{\RW{Channel access decisions determined by adopting different local measures in the timers for in a network consisting of three subsystems and two available channels.} \vspace{-0.25cm}}\label{fig:illus}	
\end{figure}

In this subsection, it is assumed that all subsystems communicate with a central scheduler which prioritizes channel access based on a measure $m_{i,j,k}$. This enables us to also consider policies which are not compatible with distributed implementation as well as the ones which might result in collisions. Fig.~\ref{fig:illus} illustrates how the choice of the prioritizing criterion affects exploration/exploitation in a small network consisting of three subsystems and two communication channels. The probability of successful transmission over each wireless link is given in table~\ref{table:t1}. Since the dynamics are identical, best performance is achieved when subsystems $1$ and $2$ transmit more frequently on Channel $1$ and Subsystem $3$ on Channel $2$.

\begin{table} [h]
\caption{Probability of successful transmission over each wireless link in the WNCS considered in Fig.~\ref{fig:illus}\vspace{-0.2cm}}
\label{table:t1}
\centering
\begin{tabular}{ |c|c|c|c| } 
	\hline
	Subsystem & 1 & 2 & 3 \\
	\hline 
	Channel 1 & $0.95$ & $0.70$ & $0.80$ \\
	\hline 
	Channel 2 & $0.81$ & $0.65$ & $0.96$ \\ 
	\hline
\end{tabular}
\end{table}
% The top plot corresponds to granting channel access based on CoIL and assuming identical quality for all wireless links. The result of ignoring CoIL and only using the UCB1 indices in the timers is depicted in the middle plot. The bottom plot illustrates the impact of using CoIL and UCB1 indices together as per \eqref{eq:timer-ucb} on the channel access decisions.
In the first scenario, we consider using CoIL as the sole priority measure without taking into account the different link qualities. To this end, we use $m_{i,j,k}={\CoIL}_{i,k}q_0$ as the priority measure, where $q_0$ is chosen as an identical constant for all links, and the higher priority subsystems choose the available channels randomly. The top plot in Fig.~\ref{fig:illus} depicts the result of adopting this scheme where, as expected, subsystems transmit over both channel equally often and the statistics are not learned thus leading to endless exploration. Next scenario concerns adopting $\hat{q}_{i,j,k}$, which is the UCB1 index determined by \eqref{eq:ucb1}, as the only priority measure. As shown in the middle plot, although the best channels are exploited in this scheme, the dynamics are ignored for the exploration/exploitation. Therefore, despite the identical dynamics, Subsystem $2$ is rarely granted channel access (starved) which could destabilize the system. Finally, we consider the effect of prioritizing with respect to CoIL as well as channel qualities by implementing the proposed policy in \eqref{eq:lambda}. As the results illustrated in the bottom plot of Fig.~\ref{fig:illus} indicate, exploration in this scheme is done with respect to CoIL while the outcome is exploited for minimizing the cost regret. More specifically, all subsystems are frequently given channel access due to their unstable dynamics. Meanwhile, Channel $1$ is allocated more often to Subsystem $1$ and Subsystem $2$ to ensure the highest probability of successful transmission, i.e., exploitation.

\begin{figure}[t]
	\centering
	\includegraphics[width=0.95\columnwidth]{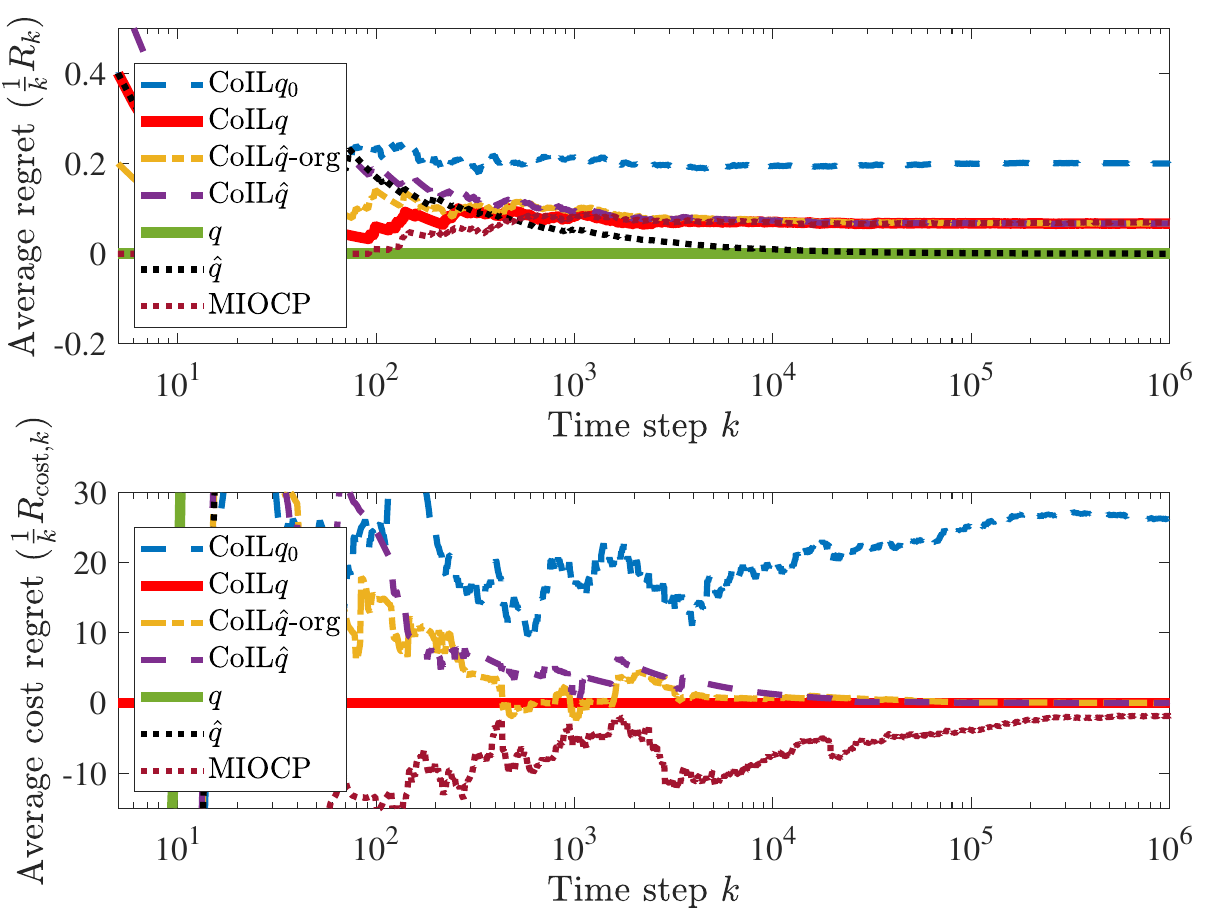}
	\vspace{-0.3cm}
	\caption{\RW{Average regret and average cost regret of three subsystems competing for two available wireless channels based on various priority measures.}\vspace{-0.25cm}}
	\label{fig:regret}	
\end{figure}
% $\CoIL q$, $q$, and MIOCP correspond to cases where the channel qualities are known and channel access is provided according to \eqref{eq:timer}, \eqref{eq:tauq}, and \eqref{eq:MIOCP} over a finite horizon, respectively. Using \eqref{eq:ucb1} in the timers or the setup in \eqref{eq:timer-ucb} are respectively represented by $\hat{q}$ and $\CoIL \hat{q}$ while $\CoIL \hat{q}\mathrm{-org}$ represents using the original UCB1 indices instead of \eqref{eq:ucb1}. $\CoIL q_0$ corresponds to granting channel access by using the timers \eqref{eq:timer} while assuming identical probability of successful packet transmission over all wireless links.
Performance of several policies in terms of average regret and cost regret is depicted in Fig.~\ref{fig:regret}. In addition to the aforementioned measures, we also consider the case of granting channel access based the solution of \eqref{eq:bestSubCh} when channel qualities are known, denoted by $q$. The solution of the MIOCP problem in \eqref{eq:MIOCP} is obtained by using the open-source nonlinear mixed integer programming (BONMIN) solver with $\kappa=5$. Moreover, we consider the impact of using the indices in \eqref{eq:ucb1}, denoted by $\CoIL \hat{q}$, instead of the originally proposed algorithm of UCB1, denoted by $\CoIL \hat{q}\mathrm{-org}$, in the timer setup \eqref{eq:timer-ucb}. As expected, the average regret while using the upper confidence bound calculated in \eqref{eq:ucb1} as the measure converges to zero, while allocating the resources without considering the channel statistics, i.e., using $\CoIL q_0$, leads to the largest average regret. On the other hand, when considering the average cost regret, the latter outperforms the scenarios where control performance is neglected. More specifically, the system is destabilized and the cost regret is unbounded for $q$ and $\hat{q}$ while $\CoIL q_0$ can stabilize the system despite its nonzero average cost regret. Moreover, the average cost regret of our proposed policy when channel qualities are unknown converges to zero fast indicating satisfactory performance. A similar trend is observed with the original UCB1 indices which shows that our proposed policy enables distributed implementation without adversely affecting the exploration/exploitation. Although using the solution of MIOCP results in lower quadratic cost than \eqref{eq:maximization} as indicated by the negative cost regret, it can only be realized in a centralized configuration and requires considerable computational resources.

%\begin{remark}
%The unstable subsystems require communication resources more frequently and since the indices are assigned with respect to CoIL, they are chosen for transmission infinitely often and hence eventually the channel statistics are learned accurately. In contrast, stable subsystems rarely transmit and thus they might learn the statistics with less accuracy. Subsequently, the unknown parameters which are essential for guaranteeing control performance are learned accurately, which implies that over time, the resources are allocated similarly to the optimal case where the channel statistics are known. Although this approach is reminiscent of those in \cite{Carpentier:2011, Carpentier:2011a}, in our case, the rewards are weighted by a function of the corresponding covariance matrix, namely CoIL, and due to the time-varying nature of CoIL providing bounds on convergence rate is still a challenging open problem.
%\end{remark}

\subsection{Distributed Implementation in Large Networks}
To evaluate the impact of the adopted learning method on performance of the timer-based mechanism, we consider three additional setups where the channel statistics are taken into account by implementing kl-UCB \cite{Cappe:2013}, kl-UCB$++$ \cite{Menard:2017}, and a Bayesian framework \cite{Farjam:2019b}. Similar to the method used for modifying UCB1, a randomly generated number is added in the exploration term of kl-UCB and kl-UCB$++$ and the resulting indices, i.e., $\hat{q}_{i,j}$, are adopted in \eqref{eq:timer-ucb} for minimizing the cost regret. The addition of the random number ensures that obtaining identical indices has Lebesgue measure zero and thus they can be used in timers for providing distributed channel access without collisions. Unlike the MAB approaches, the adopted Bayesian method is based on the assumption that the channel has memory, i.e., the packet dropouts are correlated rather than being i.i.d. random. Nevertheless, it is capable of learning the \emph{belief} of successful transmission within the timer-based framework. The cost incurred by a mechanism which ignores the channel statistics, i.e., using $\CoIL q_0$ as the measure in \eqref{eq:lambda_j}, is chosen as the benchmark for cost reduction achieved by other setups. Additionally, we consider a centralized setup which prioritizes channel access based on VoI, introduced in \cite{Molin:2015}, rather than CoIL. Since VoI is developed for resource allocation over perfect channels, we assume that one of the available channels is assigned randomly to the subsystem with the highest VoI similar to $\CoIL q_0$.

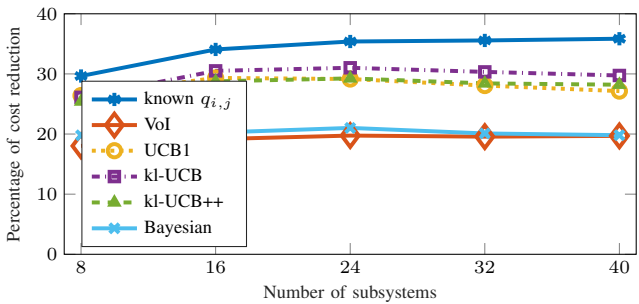
\begin{figure}[t]
\scriptsize
\centering 
\setlength\fheight{3.2cm} 
\setlength\fwidth{0.9\columnwidth}
\definecolor{mycolor1}{rgb}{0.00000,0.44700,0.74100}%
\definecolor{mycolor2}{rgb}{0.85000,0.32500,0.09800}%
\definecolor{mycolor3}{rgb}{0.92900,0.69400,0.12500}%
\definecolor{mycolor4}{rgb}{0.49400,0.18400,0.55600}%
\definecolor{mycolor5}{rgb}{0.46600,0.67400,0.18800}%
\definecolor{mycolor6}{rgb}{0.30100,0.74500,0.93300}%
\begin{tikzpicture}

\begin{axis}[%
width=0.954\fwidth,
height=\fheight,
at={(0\fwidth,0\fheight)},
scale only axis,
xmin=7,
xmax=41,
xtick={ 8, 16, 24, 32, 40},
xlabel style={font=\color{white!15!black}},
xlabel={Number of subsystems},
ymin=0,
ymax=40,
ylabel style={font=\color{white!15!black}},
ylabel={Percentage of cost reduction},
axis background/.style={fill=white},
legend style={at={(0.03,0.03)}, anchor=south west, legend cell align=left, align=left, draw=white!15!black}
]
\addplot [color=mycolor1, line width=1.5pt, mark size=2.5pt, mark=asterisk, mark options={solid, mycolor1}]
  table[row sep=crcr]{%
8	29.6509073446911\\
16	34.0696250146135\\
24	35.3897594378709\\
32	35.5659987956041\\
40	35.8505521042049\\
};
\addlegendentry{known ${q}_{i,j}$}

\addplot [color=mycolor2, line width=1.5pt, mark size=4.3pt, mark=diamond, mark options={solid, mycolor2}]
  table[row sep=crcr]{%
8	18.021926415988\\
16	19.154510139761\\
24	19.7448284646621\\
32	19.5659131202465\\
40	19.695642072102\\
};
\addlegendentry{VoI}

\addplot [color=mycolor3, dotted, line width=1.5pt, mark size=2.5pt, mark=o, mark options={solid, mycolor3}]
  table[row sep=crcr]{%
8	26.403732828582\\
16	29.3341295725663\\
24	29.1498711360989\\
32	28.0655227720396\\
40	27.1411458752696\\
};
\addlegendentry{UCB1}

\addplot [color=mycolor4, dashdotted, line width=1.5pt, mark size=1.8pt, mark=square, mark options={solid, mycolor4}]
  table[row sep=crcr]{%
8	26.1036551044723\\
16	30.5070029310062\\
24	31.0181032362565\\
32	30.3355987121197\\
40	29.7389610104929\\
};
\addlegendentry{kl-UCB}

\addplot [color=mycolor5, dashed, line width=1.5pt, mark size=1.7pt, mark=triangle, mark options={solid, mycolor5}]
  table[row sep=crcr]{%
8	25.4059160760967\\
16	28.761095671239\\
24	29.2568211390085\\
32	28.4230596314968\\
40	28.2037619722894\\
};
\addlegendentry{kl-UCB++}

\addplot [color=mycolor6, line width=1.5pt, mark size=2.5pt, mark=x, mark options={solid, mycolor6}]
  table[row sep=crcr]{%
8	19.7859601723348\\
16	20.2381068015406\\
24	21.0204769282907\\
32	20.0980076873648\\
40	19.8085024123248\\
};
\addlegendentry{Bayesian}

\end{axis}
\end{tikzpicture}%
\vspace{-0.3cm}
\caption{\RW{Reduction in the average quadratic cost achieved by adopting various timer setups compared with $m_{i,j,k}={\CoIL}_{i,k}q_0$ in WNCSs with $M=0.75 N$ channels.}\vspace{-0.3cm}} 
\label{fig:res-iid34}	
\end{figure}

Fig.~\ref{fig:res-iid34} depicts how much the average quadratic cost in \eqref{eq:cost} is reduced by the aforementioned setups compared to adopting  $\CoIL q_0$ for $N\in\{8,16,24,40\}$ and $M=0.75 N$. As expected, the best performance, i.e., lowest average cost, is achieved when the probability of successful transmission over each link is known and incorporated in \eqref{eq:timer}. This setup can reduce the incurred cost from $30\%$ to $36\%$ depending on the size of the WNCS. When the exact values of $q_{i,j}$'s are unknown, ignoring them as in VoI leads to the least amount of improvement. Nevertheless, it offers up to $20\%$ reduction in cost compared with $\CoIL q_0$ due to utilizing the measured output for prioritizing channel access rather than the statistics of the error. Using the false assumption of Markovian packet dropouts and applying the Bayesian learning method leads to better performance in smaller networks while its performance deteriorates in larger settings. When $q_{i,j}$'s are unknown \emph{a priori}, using the indexing policies for channel access results in the best performance. The results indicate that regardless of the adopted indexing policy, the setup in \eqref{eq:timer-ucb} leads to significant improvements ranging up to $30\%$. Nevertheless, utilizing the indices obtained by kl-UCB offers $1\%$ better performance compared with kl-UCB$++$ and UCB1.

% =====================================================
%
%
% Conclusions
%
%
% =====================================================
\section{Conclusion and Future Directions}\label{sec:conclusion}

\subsection{Conclusion}
In this paper, we presented a novel distributed deterministic channel access mechanism for WNCSs with imperfect (and possibly unknown) communication links. We adopted local timers for prioritizing channel access and derived the optimal timer setup for improving performance in terms of a linear quadratic cost in a distributed manner. In case of unknown channel parameters, we cast the channel access problem as a MAB and proposed a novel policy for distributed deterministic channel access. This policy utilized well-known indexing policies for estimating the success probability of channels and weighs them by a time-varying control measure, namely CoIL, which were then incorporated in timers. The simulations showed that the best performance with the timer-based mechanism is achieved when the channel parameters are known \emph{a priori}. When the parameters are unknown, however, implementing our proposed policy leads to significant improvement when compared to policies in which the channel statistics are ignored.	

\subsection{Future Directions}

Part of ongoing research is the consideration of more advanced models for the communication channels; for instance, channels with temporally correlated state variations. Another ongoing direction concerns the scenario in which flag packets have non-negligible duration which results in nonzero probability of collision between data packets. While for the deterministic case it can only result in deteriorated performance, probabilistic models can become more relevant, especially in cases where a large number of subsystems shares a limited number of channels. Furthermore, extending the proposed channel access method to WNCSs which involve subsystems with coupled dynamics poses another interesting yet challenging problem.    

%Finally, another possible direction is to extend these concepts to more general costs and to scenarios in which the cost of each subsystem (and hence priority) differs.
%This correlations can help improve the performance considerably. 

% =====================================================
%
%
% ACKNOWLEDGEMENTS
%
%
% =====================================================
%\section*{Acknowledgements}

% =====================================================
%
%
% Bibliography
%
%
% =====================================================
\bibliographystyle{IEEEtran}
\bibliography{refs_JabRefExtract}

\begin{thebibliography}{10}
\providecommand{\url}[1]{#1}
\csname url@samestyle\endcsname
\providecommand{\newblock}{\relax}
\providecommand{\bibinfo}[2]{#2}
\providecommand{\BIBentrySTDinterwordspacing}{\spaceskip=0pt\relax}
\providecommand{\BIBentryALTinterwordstretchfactor}{4}
\providecommand{\BIBentryALTinterwordspacing}{\spaceskip=\fontdimen2\font plus
\BIBentryALTinterwordstretchfactor\fontdimen3\font minus
  \fontdimen4\font\relax}
\providecommand{\BIBforeignlanguage}[2]{{%
\expandafter\ifx\csname l@#1\endcsname\relax
\typeout{** WARNING: IEEEtran.bst: No hyphenation pattern has been}%
\typeout{** loaded for the language `#1'. Using the pattern for}%
\typeout{** the default language instead.}%
\else
\language=\csname l@#1\endcsname
\fi
#2}}
\providecommand{\BIBdecl}{\relax}
\BIBdecl

\bibitem{Farjam:2019a}
T.~Farjam, T.~Charalambous, and H.~Wymeersch, ``Timer-based distributed channel
  access for control over unknown unreliable time-varying communication
  channels,'' in \emph{European Control Conference ({ECC})}, Jun. 2019.

\bibitem{Zhang:2017}
D.~Zhang, P.~Shi, Q.-G. Wang, and L.~Yu, ``Analysis and synthesis of networked
  control systems: A survey of recent advances and challenges,'' \emph{{ISA}
  Transactions}, vol.~66, pp. 376--392, Jan. 2017.

\bibitem{Schenato:2007}
L.~Schenato, B.~Sinopoli, M.~Franceschetti, K.~Poolla, and S.~S. Sastry,
  ``Foundations of control and estimation over lossy networks,''
  \emph{Proceedings of the {IEEE}}, vol.~95, no.~1, pp. 163--187, Jan. 2007.

\bibitem{Wu:2014}
J.~Wu, K.~H. Johansson, and L.~Shi, ``A stochastic online sensor scheduler for
  remote state estimation with time-out condition,'' \emph{{IEEE} Transactions
  on Automatic Control}, vol.~59, no.~11, pp. 3110--3116, Nov. 2014.

\bibitem{Wu:2013}
J.~Wu, Q.-S. Jia, K.~H. Johansson, and L.~Shi, ``Event-based sensor data
  scheduling: Trade-off between communication rate and estimation quality,''
  \emph{{IEEE} Transactions on Automatic Control}, vol.~58, no.~4, pp.
  1041--1046, Apr. 2013.

\bibitem{Nourian:2014}
M.~Nourian, A.~S. Leong, and S.~Dey, ``Optimal energy allocation for {K}alman
  filtering over packet dropping links with imperfect acknowledgments and
  energy harvesting constraints,'' \emph{{IEEE} Transactions on Automatic
  Control}, vol.~59, no.~8, pp. 2128--2143, Aug. 2014.

\bibitem{Han:2017}
D.~Han, J.~Wu, H.~Zhang, and L.~Shi, ``Optimal sensor scheduling for multiple
  linear dynamical systems,'' \emph{Automatica}, vol.~75, pp. 260--270, Jan.
  2017.

\bibitem{Quevedo:2013}
D.~E. Quevedo, A.~Ahlen, and K.~H. Johansson, ``State estimation over sensor
  networks with correlated wireless fading channels,'' \emph{{IEEE}
  Transactions on Automatic Control}, vol.~58, no.~3, pp. 581--593, Mar. 2013.

\bibitem{Huang:2007}
M.~Huang and S.~Dey, ``Stability of {K}alman filtering with {M}arkovian packet
  losses,'' \emph{Automatica}, vol.~43, no.~4, pp. 598--607, Apr. 2007.

\bibitem{Leong:2012}
A.~S. Leong and S.~Dey, ``Power allocation for error covariance minimization in
  {K}alman filtering over packet dropping links,'' in \emph{{IEEE} Conference
  on Decision and Control ({CDC})}, Dec. 2012.

\bibitem{Knorn:2017}
S.~Knorn and S.~Dey, ``Optimal energy allocation for linear control with packet
  loss under energy harvesting constraints,'' \emph{Automatica}, vol.~77, pp.
  259--267, Mar. 2017.

\bibitem{Park:2018}
P.~Park, S.~C. Ergen, C.~Fischione, C.~Lu, and K.~H. Johansson, ``Wireless
  network design for control systems: A survey,'' \emph{{IEEE} Communications
  Surveys {\&} Tutorials}, vol.~20, no.~2, pp. 978--1013, 2018.

\bibitem{Zanon:2018}
M.~Zanon, T.~Charalambous, H.~Wymeersch, and P.~Falcone, ``Optimal scheduling
  of downlink communication for a multi-agent system with a central observation
  post,'' \emph{{IEEE} Control Systems Letters}, vol.~2, no.~1, pp. 37--42,
  Jan. 2018.

\bibitem{HernandezLerma:1996}
O.~Hern{\'{a}}ndez-Lerma and J.~B. Lasserre, \emph{Discrete-Time Markov Control
  Processes: Basic Optimality Criteria}.\hskip 1em plus 0.5em minus 0.4em\relax
  Springer New York, 1996.

\bibitem{Walsh:2002}
G.~Walsh, H.~Ye, and L.~Bushnell, ``Stability analysis of networked control
  systems,'' \emph{{IEEE} Transactions on Control Systems Technology}, vol.~10,
  no.~3, pp. 438--446, May 2002.

\bibitem{Molin:2015}
A.~Molin, C.~Ramesh, H.~Esen, and K.~H. Johansson, ``Innovations-based priority
  assignment for control over {CAN}-like networks,'' in \emph{{IEEE} Conference
  on Decision and Control ({CDC})}.\hskip 1em plus 0.5em minus 0.4em\relax
  {IEEE}, Dec. 2015.

\bibitem{Molin:2019}
A.~Molin, H.~Esen, and K.~H. Johansson, ``Scheduling networked state estimators
  based on {V}alue of {I}nformation,'' \emph{Automatica}, vol. 110, p. 108578,
  Dec. 2019.

\bibitem{Charalambous:2017}
T.~Charalambous, A.~Ozcelikkale, M.~Zanon, P.~Falcone, and H.~Wymeersch, ``On
  the resource allocation problem in wireless networked control systems,'' in
  \emph{{IEEE} Conference on Decision and Control ({CDC})}, Dec. 2017.

\bibitem{Gatsis:2014}
K.~Gatsis, A.~Ribeiro, and G.~J. Pappas, ``Optimal power management in wireless
  control systems,'' \emph{{IEEE} Transactions on Automatic Control}, vol.~59,
  no.~6, pp. 1495--1510, Jun. 2014.

\bibitem{Gatsis:2015}
K.~Gatsis, M.~Pajic, A.~Ribeiro, and G.~J. Pappas, ``Opportunistic control over
  shared wireless channels,'' \emph{{IEEE} Transactions on Automatic Control},
  vol.~60, no.~12, pp. 3140--3155, Dec. 2015.

\bibitem{Leong:2017}
A.~S. Leong, D.~E. Quevedo, T.~Tanaka, S.~Dey, and A.~Ahl{\'{e}}n,
  ``Event-based transmission scheduling and {LQG} control over a packet
  dropping link,'' \emph{{IFAC}-{PapersOnLine}}, vol.~50, no.~1, pp.
  8945--8950, Jul. 2017.

\bibitem{Mamduhi:2017}
M.~H. Mamduhi, M.~Vilgelm, W.~Kellerer, and S.~Hirche, ``Prioritized contention
  resolution for random access networked control systems,'' in \emph{{IEEE}
  Conference on Decision and Control ({CDC})}, Dec. 2017.

\bibitem{Farjam:2018}
T.~Farjam, T.~Charalambous, and H.~Wymeersch, ``A timer-based distributed
  channel access mechanism in networked control systems,'' \emph{{IEEE}
  Transactions on Circuits and Systems {II}: Express Briefs}, vol.~65, no.~5,
  pp. 652--656, May 2018.

\bibitem{Recht:2019}
B.~Recht, ``A tour of reinforcement learning: The view from continuous
  control,'' \emph{Annual Review of Control, Robotics, and Autonomous Systems},
  vol.~2, no.~1, pp. 253--279, May 2019.

\bibitem{Wu:2019}
S.~Wu, X.~Ren, Q.-S. Jia, K.~H. Johansson, and L.~Shi, ``Learning optimal
  scheduling policy for remote state estimation under uncertain channel
  condition,'' \emph{IEEE Transactions on Control of Network Systems}, vol.~7,
  no.~2, pp. 579--591, Jun. 2020.

\bibitem{Gatsis:2021}
K.~Gatsis and G.~J. Pappas, ``Statistical learning for analysis of networked
  control systems over unknown channels,'' \emph{Automatica}, vol. 125, p.
  109386, Mar. 2021.

\bibitem{Wang:2020}
J.~Wang, X.~Ren, Y.~Mo, and L.~Shi, ``Whittle index policy for dynamic
  multichannel allocation in remote state estimation,'' \emph{{IEEE}
  Transactions on Automatic Control}, vol.~65, no.~2, pp. 591--603, Feb. 2020.

\bibitem{Gupta:2007}
V.~Gupta, B.~Hassibi, and R.~M. Murray, ``Optimal {LQG} control across
  packet-dropping links,'' \emph{Systems {\&} Control Letters}, vol.~56, no.~6,
  pp. 439--446, Jun. 2007.

\bibitem{Schenato:2008}
L.~Schenato, ``Optimal estimation in networked control systems subject to
  random delay and packet drop,'' \emph{{IEEE} Transactions on Automatic
  Control}, vol.~53, no.~5, pp. 1311--1317, Jun. 2008.

\bibitem{BrianD.O.Anderson:2012}
J.~B.~M. Brian D. O.~Anderson, \emph{Optimal Filtering}.\hskip 1em plus 0.5em
  minus 0.4em\relax Dover Publications, 2012.

\bibitem{Chen:1995}
G.~Chen, G.~Chen, and S.-H. Hsu, \emph{Linear stochastic control
  systems}.\hskip 1em plus 0.5em minus 0.4em\relax CRC Press, 1995.

\bibitem{Bletsas:2006}
A.~Bletsas, A.~Khisti, D.~Reed, and A.~Lippman, ``A simple cooperative
  diversity method based on network path selection,'' \emph{{IEEE} Journal on
  Selected Areas in Communications}, vol.~24, no.~3, pp. 659--672, Mar. 2006.

\bibitem{Astrom:2006}
K.~J. Astrom, \emph{Introduction to Stochastic Control Theory}.\hskip 1em plus
  0.5em minus 0.4em\relax Dover Publications Inc., 2006.

\bibitem{Kuhn:2005}
H.~W. Kuhn, ``The {H}ungarian method for the assignment problem,'' \emph{Naval
  Research Logistics}, vol.~52, no.~1, pp. 7--21, Feb. 2005.

\bibitem{Kozin:1969}
F.~Kozin, ``A survey of stability of stochastic systems,'' \emph{Automatica},
  vol.~5, no.~1, pp. 95--112, Jan. 1969.

\bibitem{Bernstein:2009}
D.~Bernstein, \emph{Matrix Mathematics: Theory, Facts, and Formulas}.\hskip 1em
  plus 0.5em minus 0.4em\relax Princeton University Press, 2009.

\bibitem{Norris:1997}
J.~R. Norris, \emph{Markov Chains}.\hskip 1em plus 0.5em minus 0.4em\relax
  Cambridge University Press, 1997.

\bibitem{Mesquita:2012}
A.~R. Mesquita, J.~P. Hespanha, and G.~N. Nair, ``Redundant data transmission
  in control/estimation over lossy networks,'' \emph{Automatica}, vol.~48,
  no.~8, pp. 1612--1620, Aug. 2012.

\bibitem{Shi:2010}
L.~Shi, M.~Epstein, and R.~M. Murray, ``Kalman filtering over a packet-dropping
  network: A probabilistic perspective,'' \emph{{IEEE} Transactions on
  Automatic Control}, vol.~55, no.~3, pp. 594--604, Mar. 2010.

\bibitem{Auer:2002}
P.~Auer, N.~Cesa-Bianchi, and P.~Fischer, ``Finite-time analysis of the
  multiarmed bandit problem,'' \emph{Machine Learning}, vol.~47, no. 2-3, pp.
  235--256, May 2002.

\bibitem{Maghsudi:2016}
S.~Maghsudi and E.~Hossain, ``Multi-armed bandits with application to 5{G}
  small cells,'' \emph{{IEEE} Wireless Communications}, vol.~23, no.~3, pp.
  64--73, Jun. 2016.

\bibitem{Pezzutto:2020}
M.~Pezzutto, F.~Tramarin, S.~Dey, and L.~Schenato, ``Adaptive transmission rate
  for {LQG} control over {W}i-{F}i: A cross-layer approach,''
  \emph{Automatica}, vol. 119, p. 109092, Sep. 2020.

\bibitem{Krikidis:2012}
I.~Krikidis, T.~Charalambous, and J.~S. Thompson, ``Buffer-aided relay
  selection for cooperative diversity systems without delay constraints,''
  \emph{{IEEE} Transactions on Wireless Communications}, vol.~11, no.~5, pp.
  1957--1967, May 2012.

\bibitem{Xu:2016}
P.~Xu, Z.~Ding, I.~Krikidis, and X.~Dai, ``Achieving optimal diversity gain in
  buffer-aided relay networks with small buffer size,'' \emph{{IEEE}
  Transactions on Vehicular Technology}, vol.~65, no.~10, pp. 8788--8794, Oct.
  2016.

\bibitem{Cappe:2013}
O.~Capp{\'{e}}, A.~Garivier, O.-A. Maillard, R.~Munos, and G.~Stoltz,
  ``{K}ullback{\textendash}{L}eibler upper confidence bounds for optimal
  sequential allocation,'' \emph{The Annals of Statistics}, vol.~41, no.~3, pp.
  1516--1541, Jun. 2013.

\bibitem{Menard:2017}
P.~M{\'e}nard and A.~Garivier, ``A minimax and asymptotically optimal algorithm
  for stochastic bandits,'' in \emph{Algorithmic Learning Theory Conference},
  Oct. 2017.

\bibitem{Farjam:2019b}
T.~Farjam, T.~Charalambous, and H.~Wymeersch, ``Timer-based distributed channel
  access in networked control systems over known and unknown
  {G}ilbert-{E}lliott channels,'' in \emph{European Control Conference
  ({ECC})}, Jun. 2019.

\end{thebibliography}

\balance

% =====================================================
%
%
% That's all folks! ;-)
%
%
% =====================================================
\end{document}